\def\TALL{\vphantom{\big|}}
\def\URL#1{$\langle${#1}$\rangle$}

\documentclass[amsmath,
nofootinbib,
preprint,prb]{revtex4}

\usepackage{graphicx}

\usepackage{pstricks}
\usepackage{pstricks-add}
\usepackage{float}
\usepackage{setspace}

\defineTColor{TRed}{magenta}%
\defineTColor{TGreen}{green}%
\defineTColor{TBlue}{red}

\def\BY#1#2{{}^{#1}_{\mbox{\scriptsize\it #2}}}
\def\BY#1#2{{}^{\mbox{\scriptsize\it (#1)}}_{\mbox{\scriptsize\it #2}}}
\def\BY#1#2{{}^{\mbox{\scriptsize\it (#1)}}_{\mbox{\scriptsize\it #2}}}
\def\BYO#1#2{{}^{\mbox{\scriptsize\it (#1)}}_{\mbox{\scriptsize\it #2}}}

\def\VEC#1{{\protect\overrightarrow{#1}}}

\def\ALICE{(Alice)}
\def\BOB{(Bob)}
\def\CAROL{(Carol)}
\def\ALICE{A}
\def\BOB{B}
\def\CAROL{C}
\def\ALICE{}
\def\BOB{'}
\def\CAROL{''}
\def\ALICE{Alice}
\def\BOB{Bob}
\def\CAROL{Carol}
\def\Alice{Alice}
\def\Bob{Bob}
\def\Carol{Carol}
\def\cspeed{{\mbox{\red \bf c}}}
\def\overc#1{ \frac{#1}{\cspeed} }
\def\overcc#1{ {#1}/{\red\cspeed^2} }
\def\overc#1{ #1 }
\def\cspeed{}
\def\overcc#1{ {#1} }
\def\VEL{\beta}

\psset{unit=.06}
\psset{framesep=0pt}

\def\pstickLINE#1{%
\psline[linecolor={#1},bordercolor=white,doublecolor=white,border=.2,doubleline=true,doublesep=.6,linestyle=solid,linewidth=0,opacity=0.6](0,0)(10,10)%
}%
\def\pstickLINELU#1{%
\psline[linecolor={#1},doublecolor=white,border=.2,doubleline=false,doublesep=.1,linestyle=dashed,dash=3 3 ,linewidth=1.0,opacity=0.06](0,0)(0,10)%
}%
\def\pstickLINELV#1{%
\psline[linecolor={#1},bordercolor=white,doublecolor=white,border=.2,doubleline=false,doublesep=.1,linestyle=dashed,dash=3 3 ,linewidth=1.0,opacity=0.06](0,0)(10,0)%
}%
\def\pstickLINEBOLD#1{%
{%
\psline[linecolor={white},bordercolor={#1},doublecolor=white,border=1.2,doubleline=false,doublesep=2,linestyle=solid,linewidth=1.5,opacity=0.06](0,0)(10,10)%
}}%
\def\pstickCONE#1{%
\psframe[linecolor={#1},dimen=middle,linestyle=dashed,linewidth=.1pt,fillstyle=none](0,0)(10,10)%
}%
\def\pstickCONETHICK#1{%
\psframe[linecolor={#1},dimen=middle,linestyle=solid,linewidth=1pt,fillstyle=none](0,0)(10,10)%
}%
\def\pstickDIAMOND#1#2{%
\psframe[linecolor={#1},dimen=middle,linestyle=solid,dotsep=.7071,linewidth=.25,fillstyle=hlines,hatchangle=30,fillcolor={#1},hatchcolor={#1},opacity=0.25](0,0)({#2})%
}%
\def\pstickSQUARE#1{%
\psframe[linecolor={#1},dimen=middle,linestyle=dotted,dotsep=.7071pt,linewidth=.7071pt,fillstyle=none](0,0)(10,10)%
}%

\def\pstickHATCH#1{%
\psframe[linecolor={#1},dimen=middle,linestyle=none,linewidth=.1pt,fillstyle=vlines,%
hatchcolor={#1},hatchsep=.500,hatchwidth=.9142,addfillstyle=vlines,opacity=0.5](0,0)(10,10)%
}%
\def\pstickHATCHLU#1{%
\psframe[linecolor=white!50,dimen=outer,linestyle=none,linewidth=0pt,fillstyle=crosshatch*,fillcolor=white!50,%
hatchcolor={#1},hatchangle=0,hatchsep=1.112,hatchwidth=.888,addfillstyle=crosshatch*,opacity=0.05](0,0)(10,10)%
}%
\def\pstickHATCHLV#1{%
\psframe[linecolor=white!50,dimen=outer,linestyle=none,linewidth=0pt,fillstyle=crosshatch*,fillcolor=white!50,%
hatchcolor={#1},hatchangle=0,hatchsep=1.112,hatchwidth=0.888,addfillstyle=crosshatch*,opacity=0.05](0,0)(10,10)%
}%
\def\pstickLIGHTLINE#1#2{%
\psline[linecolor={#1},doublecolor={#1},border=0,doubleline=false,doublesep=.6,linestyle=dashed,linewidth=1,opacity=0.6](0,0)({#2})%
}%
\def\AliceColor{magenta}
\def\BobColor{blue}
\def\CarolColor{darkgreen}
\def\DavidColor{black}
\newrgbcolor{darkgreen}{0 .5 0}

\begin{document}

\title{ Relativity on Rotated Graph Paper}

\author{Roberto B. Salgado}
\affiliation{Department of Physics and Astronomy, Bowdoin College, Brunswick, Maine 04011 USA}
\email{rsalgado@bowdoin.edu}

\begin{abstract}
We present visual calculations 
in special relativity using
spacetime diagrams drawn on graph paper that has been rotated by 45 degrees.
The rotated lines represent lightlike directions in Minkowski spacetime,
and
the boxes in the grid (called ``light-clock diamonds'')
represent units of measurement 
modeled on the ticks of an inertial observer's lightclock.
We show that
many quantitative results can be read off a spacetime diagram
by \textit{counting boxes}, using a minimal amount of algebra.
We use the Doppler Effect, in the
spirit of the Bondi $k$-calculus, to motivate the method.

\end{abstract}

\maketitle
\section{Introduction}
\label{sec:intro}

Soon after Einstein's 1905 paper\cite{Ein1905} 
on special relativity revolutionized our understanding of space and time, 
Minkowski\cite{Minkowski} 
introduced the ``spacetime diagram'' (akin to an ordinary position vs.~time graph) 
on which relativistic concepts can be encoded geometrically.
Unfortunately, %
due to the spacetime diagram's underlying non-euclidean geometry,
it may not be evident where to place tickmarks representing
units of elapsed-time and spatial-distance for 
various inertial observers in special relativity.
Without such tickmarks, spacetime diagrams may be difficult to
draw, to use for calculations, and to interpret physically.
Although one could draw the invariant hyperbolas or perform algebraic calculations
with the Lorentz Transformations,
either approach may be too sophisticated for a novice.

To help overcome these difficulties, we present a graphical method of calculation
using a (1+1)-dimensional Minkowski spacetime diagram that is
drawn on ordinary graph paper rotated by 45 degrees.  
The rotated grid will be interpreted (in Sec. \ref{sec:Radar}) 
as the radar-time light-cone coordinates\cite{DiracCoords} of an inertial observer,
where the unit squares 
(later called ``light-clock diamonds'')
mark off tickmarks for that observer's coordinate system.

Using a geometric construction, we determine an analogous grid of unit rectangles
to mark off tickmarks for another inertial observer's coordinate system.
When the relative-velocities between inertial observers are rational numbers,
calculations in special relativity can be read off the rotated grid 
by counting unit rectangles and doing simple arithmetic 
(without the explicit use of relativistic formulas).
This is the main result of this article.

The construction (to be described in detail in Sec.~\ref{sec:BobLightClock}) 
exploits the following geometrical facts%
\cite{area, Salgado, Mermin, Jacobson}
which will be motivated and interpreted physically:
(1) a rectangle in the grid [with its lightlike sides] 
(later called a ``causal diamond''\cite{causalDiamond})
has an area\cite{fn:area} 
proportional to
the square-interval%
\cite{fn:conventions} %
of its timelike-diagonal, 
and (2) a Lorentz boost transforms a rectangle in the grid
into another rectangle in the grid with 
equal area but different aspect ratio, 
stretched in one lightlike direction by the Doppler factor $k$
and compressed in the other lightlike direction by the same factor.
These geometrical facts encode properties of the invariant hyperbolas 
of Minkowski spacetime.

We begin in Sec.~\ref{sec:AliceLightClock} with an 
explanation of Alice's Light Clock and its representation on
the rotated graph paper. In Sec.~\ref{sec:UsefulTerminology},
we introduce the causal and light-clock diamond
and some additional terminology useful for describing and reasoning with spacetime diagrams.
We construct Bob's Light Clock in Sec.~\ref{sec:BobLightClock}
in two different ways,
using the Doppler Effect.
(Although our methods are in the spirit of the Bondi $k$-calculus,\cite{Bondi}
no familiarity with it is assumed.)
We apply our methods to some standard examples from special relativity
in Sec.~\ref{sec:Applications}.
In particular, in subsection~\ref{sec:Radar}, we develop a presentation
based on radar-methods, which provides 
operational definitions of physical quantities.
We conclude 
in Sec.~\ref{sec:AreaOfACausalDiamond}
with an algebraic interpretation of the construction.

With our graphical method, 
we are able to place emphasis on the physical interpretation first,
which may be followed by the development of the relativistic formulas
(to handle more general situations), if desired.
The simplicity of the method suggests that
it could be useful in an introductory course. 
(The author has used aspects of the method in such courses.)

\section{The Ticks of Alice's Light-Clock}
\label{sec:AliceLightClock}

We begin the construction by motivating the interpretation
 of the unit squares in the rotated grid.
Consider an inertial observer, Alice, at rest in her reference frame,
carrying a mirror a constant distance $D$ away. 
Alice emits a light flash (traveling with speed $c$)
which reflects off the distant mirror
and returns (at speed $c$) to her after a round-trip
elapsed time 
$%
2D/c$. If this returning light-signal is 
immediately reflected back, 
this functions like a clock, 
called the light-clock.\cite{Ein1905, Robb, Marzke, Operational, Salgado}
Since we wish to regard time as a more primitive concept than space, let us
declare that round-trip time 
to be $1 \mbox{ ``tick''}$,
so that $D=(1/2)c\rm{\ tick}=(1/2)\mbox{ ``light-tick''}$ 
[analogous to the light-year as a unit of distance].
For convenience, let us define $d=(D/c)$ so that the spatial displacement $d$ 
is also measured in ``ticks.'' Thus, $d$ measures the duration for light 
to travel the desired spatial displacement.
In these units, $c=(1\ \rm{tick})/\rm{tick}$, or simply $c=1$. 
More generally, velocities will be dimensionless and will be denoted by the symbol $\beta$.

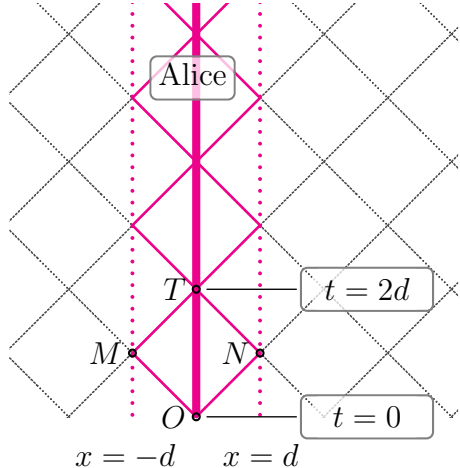
\begin{figure}[!Hht]

\def\aliceDTSCALE{\psscalebox{1 1}}
\def\aliceDXSCALE{\psscalebox{1 1}}
\def\aliceDT{\multirput[bl]{0}(0,0)(10,10){5}}%
\def\aliceDX{\multirput[ur]{-90}(50,50)(10,-10){3}}%
\def\aliceDTHATCH{\aliceDT{\aliceDTSCALE{\pstickHATCH{\AliceColor}}}}%
\def\aliceDTLINE{\aliceDT{\aliceDTSCALE{\pstickLINE{\AliceColor}}}}%
\def\aliceDTCONETHICK{\aliceDT{\aliceDTSCALE{\pstickCONETHICK{\AliceColor}}}}%
\def\aliceDXHATCH{\aliceDX{\aliceDXSCALE{\pstickHATCH{\AliceColor!75}}}}%
\def\aliceDXLINE{\aliceDX{\aliceDXSCALE{\pstickLINE{\AliceColor!75}}}}%
\def\aliceDXCONE{\aliceDX{\aliceDXSCALE{\pstickCONE{\AliceColor!75}}}}%

{%
\psset{unit=2}
\begin{pspicture}(-35,-5)(15,46)%
\psclip{\psframe[linestyle=none](-35,-0.25)(15,46)}{%
\rput[bl]{45}(0,0){%
\multirput[bl]{0}(0,-50)(0,10){20}{%
\multirput[bl]{0}(-50,0)(10,0){20}{\pstickSQUARE{black!75}}}}}%
\endpsclip%
\psclip{\psframe[linestyle=none](-35,-2)(15,46)}
{%
\rput[bl]{45}(-14.142,0)
{%
\aliceDTLINE%
\aliceDTCONETHICK%
\rput{-45}(0,0){%
\psline[linecolor=\AliceColor,bordercolor=\AliceColor,doublecolor=white,border=.2,doubleline=false,doublesep=.6,linestyle=dashed,linewidth=0.5,opacity=0.6](0,0)(0,50)%
\psline[linecolor=\AliceColor,linewidth=1.5pt,linestyle=dotted,opacity=0.6](7.07,0)(7.07,50)%
\psline[linecolor=\AliceColor,linewidth=1.5pt,linestyle=dotted,opacity=0.6](-7.07,0)(-7.07,50)%
}
\pscircle(0,0){.5}\rput[r]{-45}(0,0){$O\ $}%
\pscircle(10,10){.5}\rput[r]{-45}(10,10){$T\ $}%
\pscircle(0,10){.5}\rput[r]{-45}(0,10){$M\ $}%
\pscircle(10,0){.5}\rput[r]{-45}(10,0){$N\ $}%
\rput[b]{-45}(24,15){\rput[r](-2,10){\psframebox[linecolor=black!50,framearc=.3]{\psframebox*[framearc=.3,fillcolor=white,opacity=.75]{\parbox[c][3ex]{6ex}{\begin{tabular}{c}Alice\end{tabular}}}}}}%
\rput[b]{-45}(50,50){\rput[r](-2,10){\psframebox[linecolor=black!50,framearc=.3]{\psframebox*[framearc=.3,fillcolor=white,opacity=.75]{\begin{tabular}{c}Alice\end{tabular}}}}}%
}%
}%
\endpsclip%
\rput[t](-22,-3) {$x=-d$}%
\rput[t](-7,-3) {$x=d$}%
\rput{-0}(0,0)
{%
\psline[linecolor=black,bordercolor=white,doublecolor=white,border=0,doubleline=false,doublesep=.6,linestyle=solid,linewidth=0.15,opacity=0.6](-13,14.14)(-3,14.14)%
\rput[l](-3.8,14.14) {
\psframebox[linecolor=black!50,framearc=.3]{
\psframebox*[framearc=.3,fillcolor=white,opacity=.75]{\parbox[c][3ex]{8ex}{$t=2d$}}
}}
}
\rput{-0}(0,0)
{%
\psline[linecolor=black,bordercolor=white,doublecolor=white,border=0,doubleline=false,doublesep=.6,linestyle=solid,linewidth=0.15,opacity=0.6](-13,0)(-3,0)%
\rput[l](-3.8,0) {
\psframebox[linecolor=black!50,framearc=.3]{
\psframebox*[framearc=.3,fillcolor=white,opacity=.75]{\parbox[c][3ex]{8ex}{$t=0$}}
}}
}
\end{pspicture}%
}
\caption{
Alice's Longitudinal Light-Clock in her frame of reference. Alice emits light-signals at $O$ which reflect off her two mirrors 
(at $x=-d$ and $x=d$) to be received by Alice after an elapsed-time $\Delta t=2d$.
The resulting parallelogram $OMTN$ defines $1\mbox{\ ``tick''}$ 
of Alice's clock (so, $d=(1/2)\rm{\ tick}$),
which can be used to set up a coordinate system for Alice.
}
\label{fig:AliceLightClock}
\end{figure}

For symmetry, %
consider Alice with {\em two} such mirrors, 
one to the right (the direction in which Alice faces)
and the other to the left, 
so that they are spatially separated by $1$ tick.
Using the rotated grid, the worldlines of Alice and her mirrors (drawn as dotted lines)
and the associated light-signals are shown on the spacetime diagram in Fig.~\ref{fig:AliceLightClock}.
The parallelogram $OMTN$ will be used to represent one tick of Alice's longitudinal light-clock.
Assuming translation symmetry in spacetime, Alice can set up a 
coordinate system in which the squares in the grid 
are modeled after this parallelogram.

\section{Causal Diamonds}
\label{sec:UsefulTerminology}

\subsection{
Visualizing Proper-time with Causal Diamonds
}
We pause to introduce some terminology which will be useful for 
interpreting the diagrams physically and operationally,\cite{Minkowski,Bondi,Geroch,EllisWilliams} 
in the spacetime viewpoint.
This brief section assumes some
familiarity with the spacetime geometry of special relativity%
\cite{Minkowski,Bondi, SyngeGR, TaylorWheeler, TaylorWheelerB, Moore, Geroch, EllisWilliams, Woodhouse, Dray}
and is intended 
to give the educator an overview of the construction. 
Aspects of this section were presented to students 
after developing the construction in Sec.~\ref{sec:proofForStudents}
and then treating some examples.

The future-light-cone of an event $O$ is 
the set of events met by light-signals emitted at $O$.
Similarly, the past-light-cone of $O$ is 
the set of events whose emitted light-signals meet $O$. 
In Fig.~\ref{fig:AliceLightClock}, the future-light-cone of $O$ is the 
pair of rays from $O$ along the grid
through $OM$ and $ON$, and the past-light-cone of event $T$ is the 
pair of rays from $T$ along the grid through $TM$ and $TN$.

\begin{figure}[!Hht]

\def\aliceDTSCALE{\psscalebox{1 1}}
\def\aliceDXSCALE{\psscalebox{1 1}}
\def\aliceDT{\multirput[bl]{0}(0,0)(10,10){7}}%
\def\aliceDX{\multirput[ur]{-90}(0,0)(10,-10){6}}%
\def\aliceDTHATCH{\aliceDT{\aliceDTSCALE{\pstickHATCH{\AliceColor}}}}%
\def\aliceDTLINE{\aliceDT{\aliceDTSCALE{\pstickLINE{\AliceColor}}}}%
\def\aliceDTCONETHICK{\aliceDT{\aliceDTSCALE{\pstickCONETHICK{\AliceColor}}}}%
\def\aliceDXHATCH{\aliceDX{\aliceDXSCALE{\pstickHATCH{\AliceColor!75}}}}%
\def\aliceDXLINE{\aliceDX{\aliceDXSCALE{\pstickLINE{\AliceColor!75}}}}%
\def\aliceDXCONE{\aliceDX{\aliceDXSCALE{\pstickCONE{\AliceColor!75}}}}%

{%
\psset{unit=.85}
\begin{pspicture}(-50,0)(50,92)%
\psclip{\psframe[linestyle=none](-50,-10.25)(50,92)}{%
\rput[bl]{45}(0,0){%
\multirput[bl]{0}(0,-50)(0,10){20}{%
\multirput[bl]{0}(-50,0)(10,0){20}{\pstickSQUARE{black!75}}}}}%
\endpsclip%
\psclip{\psframe[linestyle=none](-50,-12)(50,92)}
{%
\rput[bl]{45}(-14.142,0)
{%
\aliceDTHATCH%
\aliceDXHATCH%
\aliceDTLINE%
\aliceDXLINE%
\psline[linecolor=white!25,linewidth=.3]{c-c}(10,0)(0,0)
\pscircle(0,0){.75}\rput[r]{-45}(0,0){$O\ $}%
\pscircle(10,10){.75}\rput[r]{-45}(10,10){$T\ $}%
\pscircle(0,10){.75}\rput[r]{-45}(0,10){$M\ $}%
\pscircle(10,0){.75}\rput[l]{-45}(10,0){\,$N$}%
\pscircle(10,-10){.75}\rput[c]{-45}(5,-15){$X$}%
\pscircle(80,20){.75}\rput[l]{-45}(80,20){\ $Q$}%
\rput[b]{-45}(50,50){\rput[l](7,9){\psframebox[linecolor=black!50,framearc=.3]{\psframebox*[framearc=.3,fillcolor=white,opacity=.75]{\begin{tabular}{c}Alice's\\$t$-axis\end{tabular}}}}}%
\rput[b]{-45}(35,-25){\rput[l](-8,10){\psframebox[linecolor=black!50,framearc=.3]{\psframebox*[framearc=.3,fillcolor=white,opacity=.75]{\begin{tabular}{c}Alice's\\$x$-axis\end{tabular}}}}}%
}%
}%
\endpsclip%
\end{pspicture}%
}
\caption{
Alice's Rectangular Coordinate System uses the {\it diagonals} 
of her light-clock diamonds to locate events. 
Alice's time-axis is marked by a string of light-clock diamonds 
with their timelike-diagonals along $\VEC{OT}$ on her worldline.
Alice's spatial $x$-axis is marked by a string of light-clock diamonds 
with their spacelike-diagonals along $\VEC{OX}$,
on a line of simultaneity 
according to Alice (parallel to $MN$)
through event $O$.
For example, since 
$\VEC{OQ}=5\,\VEC{OT}+3\,\VEC{OX}$,
Alice assigns event~$Q$ rectangular coordinates 
$(t\BY{\ALICE}{Q},x\BY{\ALICE}{Q})=(5,3)$.}
\label{fig:AliceCoords}
\end{figure}
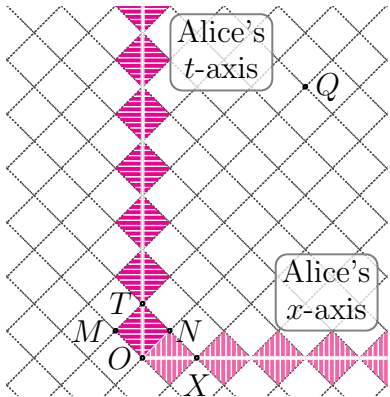

Consider now the parallelogram $OMTN$ in Fig.~\ref{fig:AliceLightClock}, 
whose sides are traced out by the light-signals of the light-clock. 
Note that diagonal $OT$ (along Alice's worldline) is future-timelike, 
and diagonal $MN$ is spacelike.
In fact,
the reflection events $M$ and $N$ are said to be 
``simultaneous according to Alice'' 
since light-signals from event $O$ to  events $M$ and $N$
are received at a common event $T$.
(Geometrically, events on the intersection of the future-light cone of $O$ and
the past-light-cone of $T$ are simultaneous according to Alice.)
Moreover, Alice regards events $M$ and $N$ as mutually simultaneous 
with her ``half-tick,'' the midpoint event of $OT$.
Thus, physically, $MN$ represents a purely-spatial direction for 
Alice when her clock reads
$t=\frac{1}{2} \rm{\ tick}$. 
This operationally defines the geometric notion that 
the spacelike-diagonal $MN$ is ``spacetime-perpendicular''\cite{fn:MinkowskiNormal}
to the timelike-diagonal $OT$.
In fact, 
Alice's spatial $x$-axis (when her clock reads $t=0 \rm{\ ticks}$)
is the ray from event $O$ that is parallel to $MN$.
See Fig.~\ref{fig:AliceCoords}.
(In the reference frame shown in Figs.~\ref{fig:AliceLightClock} and~\ref{fig:AliceCoords}, 
$MN$ happens to also be perpendicular to $OT$ in the usual Euclidean sense.)

Note, however, that the lightlike sides $OM$ and $ON$ are not 
spacetime-perpendicular%
\cite{fn:notMinkowskiPerpendicular}
to each other. Thus, $OMTN$ is neither a square nor a rectangle in Minkowskian geometry.
Instead, we will refer to parallelogram $OMTN$ as the ``causal diamond of $OT$''%
\cite{causalDiamond}
(whose interior is the intersection of the future of event $O$ 
and the past of event $T$).
In what follows, we will develop the idea that 
causal diamonds are quantitative
visualizations of the elapsed proper time along
its timelike diagonal.

\subsection{Coordinates Systems for Counting Light-Clock Diamonds}
\label{sec:coordsys}

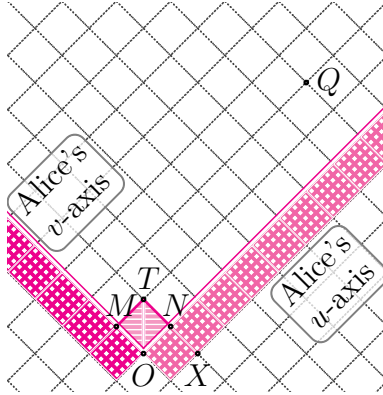
\begin{figure}[!Hht]                                              
\def\aliceDTSCALE{\psscalebox{1 1}}
\def\aliceDVSCALE{\psscalebox{1 1}}
\def\aliceDUSCALE{\psscalebox{1 1}}
\def\aliceDT{\multirput[bl]{0}(0,0)(10,10){1}}%
\def\aliceDV{\multirput[bl]{90}(0,0)(0,10){7}}%
\def\aliceDU{\multirput[br]{-90}(0,0)(10,0){9}}%
\def\aliceDTHATCH{\aliceDT{\aliceDTSCALE{\pstickHATCH{\AliceColor}}}}%
\def\aliceDTLINE{\aliceDT{\aliceDTSCALE{\pstickLINE{\AliceColor}}}}%
\def\aliceDTCONETHICK{\aliceDT{\aliceDTSCALE{\pstickCONETHICK{\AliceColor}}}}%
\def\aliceDTHATCH{\aliceDT{\aliceDTSCALE{\pstickHATCH{\AliceColor!50}}}}%
\def\aliceDVHATCH{\aliceDV{\aliceDVSCALE{\pstickHATCHLV{\AliceColor}}\psframe[linecolor=white,linestyle=solid,dotsep=.2929pt,linewidth=.3071pt](0,0)(10,10)}}%
\def\aliceDVLINE{\aliceDV{\aliceDVSCALE{\pstickLINELV{\AliceColor}}}}%
\def\aliceDVCONETHICK{\aliceDV{\aliceDVSCALE{\pstickCONETHICK{\AliceColor}}}}%
\def\aliceDUHATCH{\aliceDU{\aliceDUSCALE{\pstickHATCHLU{\AliceColor!75}}\psframe[linecolor=white,linestyle=solid,dotsep=.2929pt,linewidth=.3071pt](0,0)(10,10)}}%
\def\aliceDULINE{\aliceDU{\aliceDUSCALE{\pstickLINELU{\AliceColor!75}}}}%
\def\aliceDUCONE{\aliceDU{\aliceDUSCALE{\pstickCONE{\AliceColor!75}}}}%

\def\bobDTSCALE{\psscalebox{2 .5}}
\def\bobDT{\multirput[bl]{0}(0,0)(20,5){4}}%
\def\bobDTHATCH{\bobDT{\bobDTSCALE{\pstickHATCH{\BobColor}}}}%
\def\bobDTLINE{\bobDT{\bobDTSCALE{\pstickLINE{\BobColor}}}}%
\def\bobDTCONE{\bobDT{\bobDTSCALE{\pstickCONE{\BobColor}}}}%
{%
\psset{unit=.85}
\begin{pspicture}(-50,0)(50,92)%
\psclip{\psframe[linestyle=none](-50,-10.25)(50,92)}{%
\rput[bl]{45}(-14.142,0)
{%
}
\rput[bl]{45}(0,0){%
\multirput[bl]{0}(0,-50)(0,10){20}{%
\multirput[bl]{0}(-50,0)(10,0){20}{\pstickSQUARE{black!75}}}}}%
\endpsclip%
\psclip{\psframe[linestyle=none](-50,-12)(50,92)}
{%
\rput[bl]{45}(-14.142,0)
{%
\aliceDTHATCH%
\aliceDTLINE%
\aliceDTCONETHICK%
\rput(0.,0.){
\psline[linecolor=\AliceColor,linewidth=.6pt](0,1)(100,1)
\psline[linecolor=\AliceColor,linewidth=.6pt](1,0)(1,100)
\psline[linecolor=white,linewidth=5pt,doubleline=false,doublesep=2pt](-1,-1)(100,-1)
\psline[linecolor=white,linewidth=5pt,doubleline=false,doublesep=2pt](-1,-1)(-1,100)
}
\aliceDVHATCH%
\aliceDUHATCH%
\pscircle(0,0){.75}\rput[c]{-45}(-4,-4){$O$}%
\pscircle(10,10){.75}\rput[c]{-45}(14,14){$\ T$}%
\pscircle(0,10){.75}\rput[c]{-45}(4,14){$\ M$}%
\pscircle(10,0){.75}\rput[c]{-45}(14,4){$\ N$}%
\pscircle(10,-10){.75}\rput[c]{-45}(6,-14){$X$}%
\pscircle(80,20){.75}\rput[l]{-45}(80,20){\ $Q$}%
\rput[b]{-45}(0,50){\rput[l]{45}(5.5,-3){\psframebox[linecolor=black!50,framearc=.3]{\psframebox*[framearc=.3,fillcolor=white,opacity=.75]{\begin{tabular}{c}Alice's\\$v$-axis\end{tabular}}}}}%
\rput[b]{-45}(30,-25){\rput[l]{45}(0,5){\psframebox[linecolor=black!50,framearc=.3]{\psframebox*[framearc=.3,fillcolor=white,opacity=.75]{\begin{tabular}{c}Alice's\\$u$-axis\end{tabular}}}}}%
}%
}%
\endpsclip%
\end{pspicture}%
}
\caption{
Alice's Light-Cone Coordinate System uses the {\it edges} of her light-clock diamonds to locate events. 
Alice's $u$- and $v$-axes are marked by strings of light-clock diamonds 
with their lightlike-edges along 
$\VEC{ON}$ and $\VEC{OM}$, respectively, on the light-cone of event $O$.
Since 
$\VEC{OT}=\VEC{ON}+\VEC{OM}$ and $\VEC{OX}=\VEC{ON}-\VEC{OM}$,
we have 
$\VEC{OQ}=8\,\VEC{ON}+2\,\VEC{OM}$.
So, Alice assigns
event $Q$ the coordinates $(u\BY{\ALICE}{Q},v\BY{\ALICE}{Q})=(8,2)$.
In Sec.~\ref{sec:Radar}, we will show that 
the values of $u$ and $v$ have direct physical
interpretation as clock-readings in Alice's radar experiment to locate event $Q$.
}
\label{fig:AliceLightConeCoords}
\end{figure}

When the causal diamond of $OT$ (that is, parallelogram $OMTN$) 
is chosen as a standard to {\it mark} 
one tick of Alice's light-clock,
we will refer to it as a ``light-clock diamond of Alice,''
or simply as ``Alice's diamond.''
This provides a standard unit with which Alice can construct a coordinate system.
Since our emphasis is on \textit{counting diamonds}, we have adopted the following
choice of coordinates.
Fig.~\ref{fig:AliceCoords} shows 
Alice's rectangular coordinate system: 
$(t,x)$,
where $t$ and $x$ are 
dimensionless numbers that count the number of timelike and spacelike
\textit{diagonals} of Alice's light-clock diamond.
Fig.~\ref{fig:AliceLightConeCoords} shows 
Alice's ``light-cone coordinate'' system:\cite{DiracCoords} $(u,v)$,
where
$u$ and $v$ are dimensionless numbers that count the number 
of right- and left-pointing lightlike edges of Alice's diamond.
This coordinate system will be physically interpreted in Sec.~\ref{sec:Radar}.

Using these coordinate systems,
 an arbitrary vector $\VEC{OQ}$ can be expressed as
\begin{eqnarray}
\VEC{OQ} 
&=&
 t_Q \VEC{OT}+ x_Q\VEC{OX}  
\ =\ 
u_Q \VEC{ON} + v_Q\VEC{OM}.\label{eq:OQvector}
\end{eqnarray}
Since 
\begin{equation}
\VEC{OT}=\VEC{ON}+\VEC{OM} 
 \mbox{\qquad and\qquad} 
\VEC{OX}=\VEC{ON}-\VEC{OM},
\label{eq:BasisVectors}
\end{equation}
we find that the coordinate systems are related by
\begin{equation}
u \equiv t+x{} \mbox{\qquad and\qquad} 
v \equiv t-x{}.\label{eq:LightConeCoords}
\end{equation}
Note that the product $uv$ is equal to the square-interval
\begin{equation}
uv = t^2 -x^2,\label{eq:uv}
\end{equation}
which plays an important role in special relativity.
These relationships can be verified for the events shown
in Figs.~\ref{fig:AliceCoords} and ~\ref{fig:AliceLightConeCoords}.
Further aspects of our choice of coordinates systems will be
discussed in Sec.~\ref{sec:AreaOfACausalDiamond}.

\section{Building Bob's Light-Clock}
\label{sec:BobLightClock}

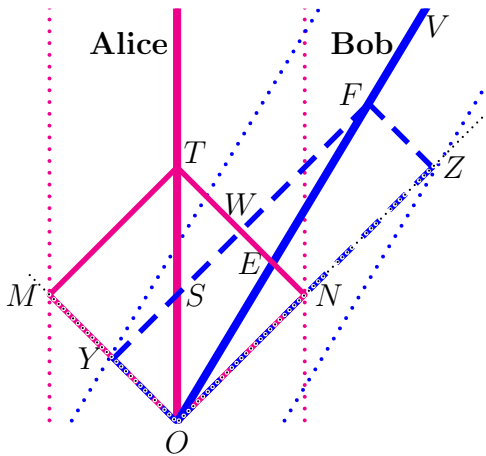
\begin{figure}[!Hh]

\def\aliceDTSCALE{\psscalebox{1 1}}
\def\aliceDXSCALE{\psscalebox{1 1}}
\def\aliceDT{\multirput[bl]{0}(0,0)(10,10){5}}%
\def\aliceDX{\multirput[ur]{-90}(50,50)(10,-10){3}}%
\def\aliceDTHATCH{\aliceDT{\aliceDTSCALE{\pstickHATCH{\AliceColor}}}}%
\def\aliceDTLINE{\aliceDT{\aliceDTSCALE{\pstickLINE{\AliceColor}}}}%
\def\aliceDTCONETHICK{\aliceDT{\aliceDTSCALE{\pstickCONETHICK{\AliceColor}}}}%
\def\aliceDXHATCH{\aliceDX{\aliceDXSCALE{\pstickHATCH{\AliceColor!75}}}}%
\def\aliceDXLINE{\aliceDX{\aliceDXSCALE{\pstickLINE{\AliceColor!75}}}}%
\def\aliceDXCONE{\aliceDX{\aliceDXSCALE{\pstickCONE{\AliceColor!75}}}}%

{%
\psset{unit=2}
\begin{pspicture}(-25,-3)(25,45)%
\rput[l](-14.14,0)
{%
\rput[b](0,42){\rput[r](-0,0){\psframebox*[framearc=.3,fillcolor=white,opacity=.75,framesep=1]{\bf Alice}}}%
\rput[b](17,42){\rput[l](0,0){\psframebox*[framearc=.3,fillcolor=white,opacity=.75]{\bf Bob}}}%
\rput[t](0,-1) {$O$}%
\rput[l](0.75, 30) {$T$}%
\rput[r](21.71,36.35) {$F$ }%
\rput[b](7.07,21.71) {$W\TALL$}%
\rput[l](-.25,14.14) { $S$}%
\rput[r](10.61,17.675) {$E$ }%
\rput[t](28.28,45.13) { $V$}%
\rput[r](-14.14,14.14) {$M$ }%
\rput[l](14.14,14.14) { $N$}%
\rput[r](-7.07,7.07) {$Y$ }%
\rput[l](28.28,28.28) { $Z$}%
}
\psclip{\psframe[linestyle=none](-30,-0.25)(20,46)}
{%
\rput[l](-14.142,0)
{%
\psline[linecolor=\AliceColor,linewidth=1.5pt,linestyle=dotted,opacity=0.6](14.14,0)(14.14,50)%
\psline[linecolor=\AliceColor,linewidth=1.5pt,linestyle=dotted,opacity=0.6](-14.14,0)(-14.14,50)%
\psline[linecolor=\BobColor,linewidth=1.5pt,linestyle=dotted,opacity=0.6](-11.785,0)(30.641,70.710)%
\psline[linecolor=\BobColor,linewidth=1.5pt,linestyle=dotted,opacity=0.6](11.785,0)(54.211,70.710)%
\psline[linecolor=\AliceColor,bordercolor=\AliceColor,doublecolor=white,border=.2,
doubleline=false,doublesep=.6,linestyle=dashed,linewidth=0.5,opacity=0.6](0,0)(0,128.284)%
\psline[linecolor=\BobColor,bordercolor=\BobColor,doublecolor=white,border=.2,
doubleline=false,doublesep=.6,linestyle=dashed,linewidth=0.5,opacity=0.6](0,0)(42.426,70.710) %
}
\rput[l]{45}(-14.142,0)
{
\psline[linecolor=\AliceColor,linewidth=.6]{c-c}(0,0)(0,20)
\psline[linecolor=\AliceColor ,linewidth=.6]{c-c}(0,20)(20,20)
\psline[linecolor=\AliceColor ,linewidth=.6]{c-c}(20,20)(20,0)
\psline[linecolor=\AliceColor ,linewidth=.6]{c-c}(20,0)(0,0)
\psline[linecolor=\BobColor,linestyle=dashed,dash= 2.5 1.5 ,linewidth=.6]{c-c}(0,0)(0,10)
\psline[linecolor=\BobColor,linestyle=dashed,dash= 2.5 1.5 ,linewidth=.6]{c-c}(0,10)(40,10)
\psline[linecolor=\BobColor,linestyle=dashed,dash= 2.5 1.5 ,linewidth=.6]{c-c}(40,10)(40,0)
\psline[linecolor=\BobColor,linestyle=dashed,dash= 2.5 1.5 ,linewidth=.6]{c-c}(40,0)(0,0)
}
}%
\endpsclip%
\rput[l](-14.14,0)%
{%
\psclip{\psframe[linestyle=none](-16.86,-0.25)(34.14,46)}%
{%
\psset{algebraic=true, showpoints=false}%
\psplot[VarStep=true, VarStepEpsilon=.001, linecolor=white, dotsep=.25, linestyle=dotted, linewidth=.5]{-50}{50}{x}%
\psplot[VarStep=true, VarStepEpsilon=.001, linecolor=black, dotsep=.5, linestyle=dotted, linewidth=.25]{-50}{50}{x}%
\psplot[VarStep=true, VarStepEpsilon=.001, linecolor=white, dotsep=.25, linestyle=dotted, linewidth=.5]{-50}{50}{-x}%
\psplot[VarStep=true, VarStepEpsilon=.001, linecolor=black, dotsep=.5, linestyle=dotted, linewidth=.25]{-50}{50}{-x}%
}%
\endpsclip%
}%
\end{pspicture}%
}
\caption{
Bob's Longitudinal Light-Clock in Alice's frame of reference.
Bob, traveling with velocity $(3/5)\cspeed$ according to Alice,
emits light-signals at $O$ which reflect off his two mirrors 
to be received by Bob after an elapsed-time of $1\rm\ tick$ on his clock.
The resulting parallelogram $OYFZ$ with timelike-diagonal $OF$ would
define Bob's light-clock diamond,
which can be used to set up a coordinate system for Bob.
So, given parallelogram $OMTN$ and $OV$, we wish to determine 
the event $F$ on $OV$ 
such that $t\BY{\ALICE}{T}=t\BY{\BOB}{F}$.
}
\label{fig:BobLightClock}
\end{figure}

Now, consider another inertial observer, Bob, 
moving with velocity $\VEL=(3/5)\cspeed$ (for numerical convenience) 
according to Alice. 
Refer to Fig.~\ref{fig:BobLightClock}.
Let $OV$ refer to Bob's worldline and 
let event $F$ on $OV$ mark off one tick from Bob's clock since event $O$.
Given $\VEL$, we use Alice's ticks to determine the line $OV$.
The problem is to locate event $F$ on~$OV$.

In accordance with the Relativity Principle and the Speed of Light Principle,\cite{Ein1905} 
we expect that the first tick of Bob's identically-constructed longitudinal light-clock will be
drawn as the causal diamond of $OF$. 
Starting from the causal diamond of $OT$ (one tick of Alice's light-clock), 
we will now
construct the causal diamond of $OF$ (one tick of Bob's light-clock).
The key result we will deduce is that 
\textit{the areas of these diamonds are equal.}

\subsection{A geometrical proof}
\label{sec:proofForEducators}

We will use the Doppler Effect 
(in the spirit of the Bondi $k$-calculus\cite{Bondi})
to elaborate on a geometric argument from Mermin.\cite{Mermin}
Our goal is to highlight the role of 
physical laws in the construction.
Since this argument may be too abstract for students, 
we will present another demonstration in Sec.~B %
(which I have used with students).
Some further aspects of the construction are treated algebraically 
in Sec.~\ref{sec:AreaOfACausalDiamond}.
(As mentioned earlier in the introduction,
no familiarity with the Bondi $k$-calculus %
is assumed.)

Refer again to Fig.~\ref{fig:BobLightClock}.
Start with the causal diamond of $OT$ and Bob's worldline $OV$.
Note the intersection events: 
the given event $O$ and the event $E$ on $T$'s past-light-cone. 
Interpret 
these events $O$ and $E$ as Bob emitting two light-signals
which are received by Alice at events $O$ and $T$, respectively.
Along segment $OE$, let $t\BY{\BOB}{E}$ represent the period between emissions according to Bob's clock, and
along segment $OT$, let $t\BY{\BOB}{T}$ represent the period between receptions according to Alice's clock.
These periods (in triangle $OET$) are related by
\begin{equation}
t\BY{\ALICE}{T} = k\BY{\ALICE}{\Bob} t\BY{\BOB}{E},
\label{eq:TkE}
\end{equation}
where $k\BY{\ALICE}{\Bob}$ is a proportionality factor determined by Alice
that depends on the relative speed $\beta$, but is independent of $t\BY{\ALICE}{T}$.
In passing, we mention that this factor, which we call the Bondi-Doppler $k$-factor,
is equal to the familiar Doppler factor
\begin{eqnarray}
k&=\sqrt{(1+\VEL)/(1-\VEL)},\label{eq:BondiKfactor}
\end{eqnarray}
as we show in Sec.~\ref{sec:Radar} (Eq.~\ref{eq:kvel}). We do not need this explicit expression in this section.

With an arbitrary event $F$ along $OV$,
construct the causal diamond of $OF$ and the 
corresponding intersection events
$O$ and $S$ on Alice's worldline.
Similarly, we have (in triangle $OSF$)
\begin{equation}
t\BY{\BOB}{F} = k\BY{\BOB}{\Alice} t\BY{\ALICE}{S},
\label{eq:FkS}
\end{equation}
where $k\BY{\BOB}{\Alice}$ is the analogous factor determined by Bob.
By the Relativity Principle, since Alice and Bob are conducting 
identical experiments on each other, we must have 
\begin{equation}
k\BY{\ALICE}{\Bob}=k\BY{\BOB}{\Alice},
\label{eq:equalKs}
\end{equation}
which we will now call $k_{\Alice,\Bob}$, or simply $k$ when there is no ambiguity. 

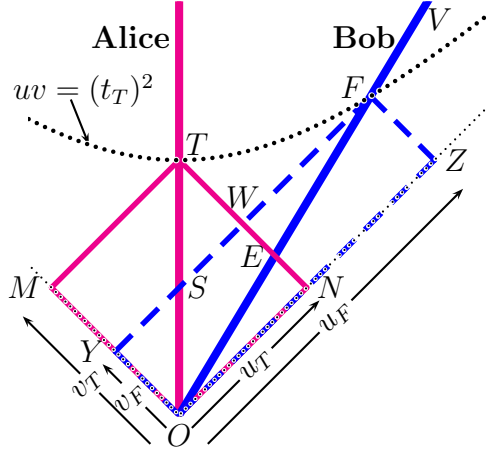
\begin{figure}[!Hht]

\def\aliceDTSCALE{\psscalebox{1 1}}
\def\aliceDXSCALE{\psscalebox{1 1}}
\def\aliceDT{\multirput[bl]{0}(0,0)(10,10){5}}%
\def\aliceDX{\multirput[ur]{-90}(50,50)(10,-10){3}}%
\def\aliceDTHATCH{\aliceDT{\aliceDTSCALE{\pstickHATCH{\AliceColor}}}}%
\def\aliceDTLINE{\aliceDT{\aliceDTSCALE{\pstickLINE{\AliceColor}}}}%
\def\aliceDTCONETHICK{\aliceDT{\aliceDTSCALE{\pstickCONETHICK{\AliceColor}}}}%
\def\aliceDXHATCH{\aliceDX{\aliceDXSCALE{\pstickHATCH{\AliceColor!75}}}}%
\def\aliceDXLINE{\aliceDX{\aliceDXSCALE{\pstickLINE{\AliceColor!75}}}}%
\def\aliceDXCONE{\aliceDX{\aliceDXSCALE{\pstickCONE{\AliceColor!75}}}}%

{%
\psset{unit=2}
\begin{pspicture}(-25,-3)(25,55)%
\rput[l](-14.14,0)
{%
\rput[b](0,42){\rput[r](-0,0){\psframebox*[framearc=.3,fillcolor=white,opacity=.75,framesep=1]{\bf Alice}}}%
\rput[b](17,42){\rput[l](0,0){\psframebox*[framearc=.3,fillcolor=white,opacity=.75]{\bf Bob}}}%
\rput[t](0,-1) {$O$}%
\rput[l](0.75, 30) {$T$}%
\rput[r](21.71,36.35) {$F$ }%
\rput[b](7.07,21.71) {$W\TALL$}%
\rput[l](-.25,14.14) { $S$}%
\rput[r](10.61,17.675) {$E$ }%
\rput[t](28.28,45.13) { $V$}%
\rput[r](-14.14,14.14) {$M$ }%
\rput[l](14.14,14.14) { $N$}%
\rput[r](-7.07,7.07) {$Y$ }%
\rput[l](28.28,28.28) { $Z$}%
}
\psclip{\psframe[linestyle=none](-30,-0.25)(20,46)}
{%
\rput[l](-14.142,0)
{%
\psline[linecolor=\AliceColor,bordercolor=\AliceColor,doublecolor=white,border=.2,
doubleline=false,doublesep=.6,linestyle=dashed,linewidth=0.5,opacity=0.6](0,0)(0,128.284)%
\psline[linecolor=\BobColor,bordercolor=\BobColor,doublecolor=white,border=.2,
doubleline=false,doublesep=.6,linestyle=dashed,linewidth=0.5,opacity=0.6](0,0)(42.426,70.710) %
}
\rput[l]{45}(-14.142,0)
{
\psline[linecolor=\AliceColor,linewidth=.6]{c-c}(0,0)(0,20)
\psline[linecolor=\AliceColor ,linewidth=.6]{c-c}(0,20)(20,20)
\psline[linecolor=\AliceColor ,linewidth=.6]{c-c}(20,20)(20,0)
\psline[linecolor=\AliceColor ,linewidth=.6]{c-c}(20,0)(0,0)
\psline[linecolor=\BobColor,linestyle=dashed,dash= 2.5 1.5 ,linewidth=.6]{c-c}(0,0)(0,10)
\psline[linecolor=\BobColor,linestyle=dashed,dash= 2.5 1.5 ,linewidth=.6]{c-c}(0,10)(40,10)
\psline[linecolor=\BobColor,linestyle=dashed,dash= 2.5 1.5 ,linewidth=.6]{c-c}(40,10)(40,0)
\psline[linecolor=\BobColor,linestyle=dashed,dash= 2.5 1.5 ,linewidth=.6]{c-c}(40,0)(0,0)
}
}%
\endpsclip%
\rput[l]{45}(-14.142,0)
{
\psline{->}(-4.5,0)(-4.5,20)
\psframe*[linecolor=white,fillcolor=white](-6,8)(-2.75,11)
\rput[c](-4.75,9.5) {$v_T$}%
\psline{->}(0,-2)(20,-2)
\psframe*[linecolor=white,fillcolor=white](7.5,-4)(12.5,-0.75)
\rput[b](10,-3.25) {$u_T$}%
\psline{->}(-2,0)(-2,10)
\psframe*[linecolor=white,fillcolor=white](-3,3.25)(-1,7)
\rput[c](-2.5,5) {$v_F$}%
\psline{->}(0,-4.5)(40,-4.5)
\psframe*[linecolor=white,fillcolor=white](17.5,-6)(22.5,-3.5)
\rput[b](20,-5.75) {\ $u_F$\ }%
}%
\rput[l](-14.14,0)%
{%
\rput[c](-10,36) {$uv=(t_{T})^2$\ }%
\psline{->}(-12,35)(-10,30)
\psclip{\psframe[linestyle=none](-16.86,-0.25)(34.14,46)}%
{%
\psset{algebraic=true, showpoints=false}%
\psplot[VarStep=true, VarStepEpsilon=.001, linecolor=white, dotsep=.25, linestyle=dotted, linewidth=.5]{-50}{50}{x}%
\psplot[VarStep=true, VarStepEpsilon=.001, linecolor=black, dotsep=.5, linestyle=dotted, linewidth=.25]{-50}{50}{x}%
\psplot[VarStep=true, VarStepEpsilon=.001, linecolor=white, dotsep=.25, linestyle=dotted, linewidth=.5]{-50}{50}{-x}%
\psplot[VarStep=true, VarStepEpsilon=.001, linecolor=black, dotsep=.5, linestyle=dotted, linewidth=.25]{-50}{50}{-x}%
\psplot[VarStep=true, VarStepEpsilon=.001, linecolor=white, dotsep=.25, linewidth=.75, linestyle=dotted]{-50}{50}{2*14.142*sqrt(1+(x/(2*14.142))^2)}%
\psplot[VarStep=true, VarStepEpsilon=.001, linecolor=black, dotsep=.5, linewidth=.5, linestyle=dotted]{-50}{50}{2*14.142*sqrt(1+(x/(2*14.142))^2)}%
}%
\endpsclip%
}%
\end{pspicture}%
}
\caption{
Calibrating Bob's Longitudinal Light-Clock. 
(A short proof based on Mermin.\cite{Mermin})  
Interpreting triangle $OSF$ as a Doppler Effect
(the period between reception events, $O$ and $F$, 
is proportional to the period between emission events, $O$ and $S$),
we have $t\BY{\BOB}{F}=k\, t\BY{\ALICE}{S}$,
where $k$ is the Bondi-Doppler factor.
Similarly, from triangle $OET$,
we have $t\BY{\ALICE}{T}=k\, t\BY{\BOB}{E}$,
with the same $k$-factor 
(in accordance with the Relativity Principle).
By similar triangles, 
$t\BY{\ALICE}{T}/t\BY{\ALICE}{S}=v_T/v_F$ and 
$t\BY{\BOB}{F}/t\BY{\BOB}{E}=u_F/u_T$.
When $t\BY{\ALICE}{T}=t\BY{\BOB}{F}$, 
it follows that $u_T v_T=u_F v_F$.  
Geometrically, this means 
the required parallelograms $OMTN$ and $OYFZ$ have equal areas, and 
that events $T$ and $F$ lie on the hyperbola $uv=(t\BY{\ALICE}{T})^2$ with asymptotes along $O$'s light cone.
}
\label{fig:BobLightClockProof}
\end{figure}

Refer now to Fig.~\ref{fig:BobLightClockProof}.
Following Mermin,\cite{Mermin} let us describe the sides of the causal diamonds 
with Alice's light-cone coordinates $v_T$, $v_F$, $u_T$, and $u_F$ so that we have the 
``areas''%
\cite{fn:areaCalc}
$u_T v_T$ and $u_F v_F$ for Alice's and Bob's causal diamonds.
By similar triangles, we have 
\begin{equation}
\frac{t\BY{\ALICE}{T}}{t\BY{\ALICE}{S}}=\frac{v_T}{v_F}
\qquad \mbox{and}\qquad
\frac{t\BY{\BOB}{F}}{t\BY{\BOB}{E}}=\frac{u_F}{u_T}.
\label{eq:similarity}
\end{equation}

Forming the ratio $(OF)/(OT)$, we use
Eq.~(\ref{eq:TkE}) (Doppler) 
with Eq.~(\ref{eq:equalKs}) (Relativity)
and Eq.~(\ref{eq:similarity})
(similiar triangles) to obtain
\begin{equation}
\frac{t\BY{\BOB}{F}}{t\BY{\ALICE}{T}}
=\frac{t\BY{\BOB}{F}}{k\ t\BY{\BOB}{E}}
=\frac{u_F}{k\ u_T}.
\label{eq:uratio}
\end{equation}
Similarly, using
Eq.~(\ref{eq:FkS}) with Eq.~(\ref{eq:equalKs})
and Eq.~(\ref{eq:similarity})
we have
\begin{equation}
\frac{t\BY{\BOB}{F}}{t\BY{\ALICE}{T}}
=\frac{k\ t\BY{\ALICE}{S}}{t\BY{\ALICE}{T}}
=\frac{k\ v_F}{v_T},
\label{eq:vratio}
\end{equation}

Upon multiplication of Eqs.~(\ref{eq:uratio}) and~(\ref{eq:vratio}),
we eliminate the $k$-factors to obtain
\begin{equation}
\left(\frac{t\BY{\BOB}{F}}{t\BY{\ALICE}{T}}\right)^2
=\frac{u_F\ v_F}{u_T\ v_T},
\end{equation}
where the right-hand side is the ratio of areas of 
two causal diamonds. This means that,
\textit{in units of the area of Alice's light-clock diamond,
the area of the causal diamond of $OF$ 
is equal to the square-interval of segment $OF$.}
To locate the event $F$ that marks Bob's first tick, 
we require
$t\BY{\BOB}{F}/t\BY{\ALICE}{T}=1$.
Thus, event $F$ is located at the opposite
corner of a causal diamond with diagonal along $OV$
and with area equal to that of Alice's light-clock diamond (along $OT$).

Now that we have constructed Bob's tick, 
three relativistic effects are immediately evident from the 
spacetime diagram (Fig.~\ref{fig:BobLightClockProof}).
Using Alice's ticks, we find that the elapsed-time for Bob's tick $OF$ 
is larger than that of her own tick $OT$. 
This is the \textit{time-dilation effect}.  
Although Alice's reflection events are simultaneous according to Alice, 
Bob's reflection events are not simultaneous according to Alice.
This is the \textit{relativity of simultaneity}.
Upon constructing the worldlines of the mirrors 
for Bob's light-clock (as we did in Fig.~\ref{fig:BobLightClock}),
Alice determines that the length of Bob's light-clock
is shorter than her identically constructed light-clock.
This is the \textit{length-contraction effect}.

Note that when $t\BY{\BOB}{F}/t\BY{\ALICE}{T}=1$,
Eqs.~(\ref{eq:uratio}) and~(\ref{eq:vratio}) can be rewritten as
\begin{eqnarray}
u_F&=&k\ u_T \label{eq:uTransform}\\
v_F&=&\frac{1}{k}\ v_T, \label{eq:vTransform}
\end{eqnarray}
which represent an area-preserving transformation of 
Alice's light-clock diamond
into Bob's light-clock diamond,
 by stretching along the $u$-direction by a factor $k$
and shrinking along the $v$-direction by a factor $k$---that is,
an aspect ratio $u_F/v_F=k^2$.
Physically, this preserves the speed of light
(since the slopes of light-rays are unchanged)
and the square-interval of the diagonal of the light-clock diamond
(since $u_F v_F=u_T v_T$).
This is an \textit{active Lorentz boost transformation}, 
expressed in a basis of its lightlike eigenvectors,
with eigenvalues $k$ and $1/k$. To recover the standard
rectangular form, write
half of the sum and half of the difference
of Eqs.~(\ref{eq:uTransform}) and~(\ref{eq:vTransform}), then
use Eq.~(\ref{eq:LightConeCoords}) and Eq.~(\ref{eq:BondiKfactor}). 
(See Eq.~(\ref{eq:BoostFormula0}) in the Appendix
for a similar calculation for the passive Lorentz Boost 
coordinate transformation.)

\subsection{A construction for use with students}
\label{sec:proofForStudents}

In this section, we construct Bob's ticks using an alternate method  
that has been used with students. Our story uses radar-experiments 
and the Doppler Effect (in the spirit of the Bondi $k$-calculus\cite{Bondi} method)
framed in the context of television transmissions sent and received by Alice and Bob.%
\cite{Calder}

With a sheet of rotated graph paper, we draw Alice's worldline and her 
light-clock diamonds.
(See Fig.~3.) While the computer-generated shading is useful for visualization,
it is not a necessary feature for visual calculations.
Using Alice's 
light-clock diamonds,
we draw Bob's worldline along $OJ$
with velocity $\VEL=(3/5)\cspeed$. We construct two segments, 
$OH$ along Alice's worldline and $HJ$ where $H$ and $J$ 
are simultaneous according to Alice,
such that $(HJ)/(OH)=(3/5)\cspeed$.
We now begin the construction of Bob's 
light-clock diamonds.

\def\aliceDTSCALE{\psscalebox{1 1}}
\def\aliceDXSCALE{\psscalebox{1 1}}
\def\aliceDT{\multirput[bl]{0}(0,0)(10,10){13}}%
\def\aliceDX{\multirput[ur]{-90}(100,100)(10,-10){6}}%
\def\aliceDTHATCH{\aliceDT{\aliceDTSCALE{\pstickHATCH{\AliceColor}}}}%
\def\aliceDTLINE{\aliceDT{\aliceDTSCALE{\pstickLINE{\AliceColor}}}}%
\def\aliceDTCONE{\aliceDT{\aliceDTSCALE{\pstickCONE{\AliceColor}}}}%
\def\aliceDXHATCH{\aliceDX{\aliceDXSCALE{\pstickHATCH{\AliceColor!75}}}}%
\def\aliceDXLINE{\aliceDX{\aliceDXSCALE{\pstickLINE{\AliceColor!75}}}}%
\def\aliceDXCONE{\aliceDX{\aliceDXSCALE{\pstickCONE{\AliceColor!75}}}}%

\def\bobDTSCALE{\psscalebox{2 .5}}
\def\bobDXSCALE{\psscalebox{.5 2}}
\def\bobDT{\multirput[bl]{0}(0,0)(20,5){10}}%
\def\bobDTA{\multirput[bl]{0}(0,0)(20,5){2}}%
\def\bobDX{\multirput[ur]{90}(200,50)(-20,5){6}}%
\def\bobDTHATCH{\bobDTA{\bobDTSCALE{\pstickHATCH{\BobColor}}}}%
\def\bobDTLINE{\bobDT{\bobDTSCALE{\pstickLINE{\BobColor}}}}%
\def\bobDTCONE{\bobDT{\bobDTSCALE{\pstickCONE{\BobColor}}}}%
\def\bobDXHATCH{\bobDX{\bobDXSCALE{\pstickHATCH{\BobColor!75}}}}%
\def\bobDXLINE{\bobDX{\bobDXSCALE{\pstickLINE{\BobColor!75}}}}%
\def\bobDXCONE{\bobDX{\bobDXSCALE{\pstickCONE{\BobColor!75}}}}%

\begin{figure}[!Hht]
{%
\psset{unit=.85}
\begin{pspicture}(-35,0)(100,184)%
\psclip{\psframe[linestyle=none](-35,-0.25)(100,184)}{%
\rput[bl]{45}(0,0){%
\multirput[bl]{0}(0,-80)(0,10){25}{%
\multirput[bl]{0}(-50,0)(10,0){25}{\pstickSQUARE{black!75}}}}}%
\endpsclip%
\rput[bl]{45}(-14.142,0)
{%
\aliceDTHATCH%
\aliceDXHATCH%
\rput{0}(80,20){\bobDTHATCH}%
\aliceDTLINE%
\aliceDXLINE%
\bobDTLINE%
\aliceDTCONE%
\aliceDXCONE%
\bobDTCONE%
\pscircle(0,0){1}\rput[r]{-45}(0,0){$O\ $}%
\pscircle(100,100){1}\rput[r]{-45}(100,100){$H\ $}%
\pscircle(160,40){1}\rput[l]{-45}(160,40){$\ J$}%
\rput[b]{-45}(174,78){\rput[t](2,-1){\psframebox*[boxsep=true,framesep=-1pt,framearc=.3,fillcolor=white,opacity=.75]{\begin{tabular}{c}$\VEL\BY{\ALICE}{\Bob}=\displaystyle\frac{3}{5}\cspeed$\end{tabular}}}}%
\rput[b]{-45}(102,102){\rput[r](-5,10){\psframebox*[framearc=.3,fillcolor=white,opacity=.75,framesep=1]{\bf Alice}}}%
\rput[b]{-45}(149,54){\rput[l](6,10){\psframebox*[framearc=.3,fillcolor=white,opacity=.75]{\bf Bob}}}%
\rput[bl]{0}(20,20){\pstickLIGHTLINE{black}{60,00}}%
\rput[bl]{0}(80,20){\pstickLIGHTLINE{black}{0,60}}%
\rput[bl]{0}(30,30){\pstickLIGHTLINE{black}{90,00}}%
\rput[bl]{0}(120,30){\pstickLIGHTLINE{black}{0,90}}%
\pscircle(120,120){1}\rput[r]{-45}(120,120){$r_3\ $}%
\pscircle(120,30){1}\rput[l]{-45}(120,30){$\ B_3$}%
\pscircle(80,80){1}\rput[r]{-45}(80,80){$r_2\,$ }%
\pscircle(80,20){1}\rput[l]{-45}(80,20){$\ B_2$}%
\pscircle(30,30){1}\rput[r]{-45}(30,30){$e_3\ $}%
\pscircle(20,20){1}\rput[r]{-45}(20,20){$e_2\ $}%
\psline[linecolor=\BobColor,linewidth=.6]{c-c}(80,20)(0,0)
\psline[linecolor=\BobColor,linewidth=.6]{c-c}(208,52)(120,30)
}%
\end{pspicture}%
}
\caption{\label{fig:BobLightClockStudents}
Calibrating Bob's Longitudinal Light-Clock. (Motivation for students.) 
}
\end{figure}
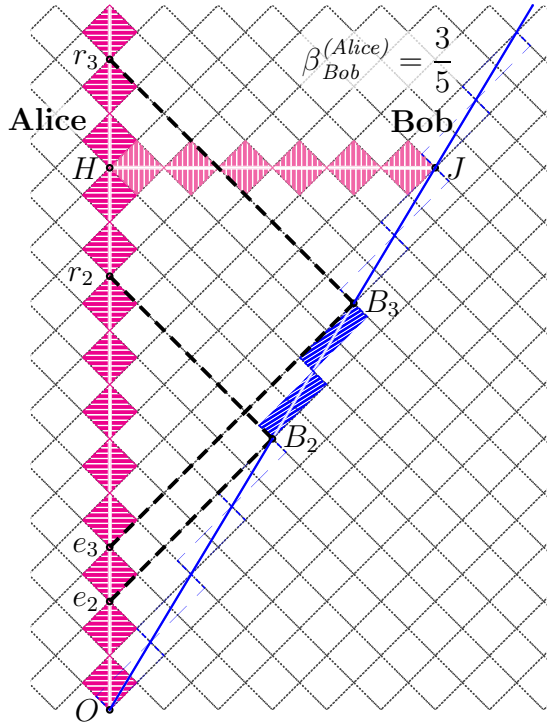

Alice sends to Bob two successive light-signals, the second signal sent one tick after the first.
We interpret these as broadcasts marking the start and end of a one-hour 
television program produced by Alice. Due to the finiteness of the speed of light,
Bob receives the first signal after a delay. However, since Bob is receding from Alice,
Bob receives the second signal after $k$ of his ticks, 
where $k$ is a proportionality constant to be determined. 
That is, 
\begin{equation}
\label{eq:BondiK}
\Delta t\BY{\BOB}{receptions}=k \Delta t\BY{\ALICE}{emissions}.
\end{equation}
(In the previous section, the constant $k$ was called $k\BY{\BOB}{\Alice}$.)
It is easy to see that $k=1$ for a distant observer at rest according to Alice, but $k>1$
for a distant inertial observer who is receding from Alice. 
Thus, Bob would take $k$ of his hours to 
watch
\cite{fn:decodeTransmissions}
in slow-motion a one-hour program by Alice.

In order to determine $k$, we arrange to have Bob immediately rebroadcast the received signals from Alice.
By the Relativity Principle, Alice must receive the delayed broadcast from Bob at the same slowed rate.
Alice would take $k$ of her hours to watch 
in slow-motion a one-hour program by Bob.
So, 
\begin{eqnarray}
\Delta t\BY{\ALICE}{receptions}
&=&k \Delta t\BY{\BOB}{emissions}\nonumber\\
&=&k (\Delta t\BY{\BOB}{receptions})
=k(k \Delta t\BY{\ALICE}{emissions})
.
\end{eqnarray}
Thus, Alice would take $k^2$ of her hours to watch 
in very slow-motion her originally broadcasted one-hour program.
By counting 
Alice's light-clock diamonds
off the spacetime diagram, 
one can determine, for $\VEL=(3/5)\cspeed$, the corresponding value of $k^2$:
\begin{equation}
\label{eq:Ksq}
k^2=\frac{\Delta t\BY{\ALICE}{receptions}}{\Delta t\BY{\ALICE}{emissions}}=
\frac{ t\BY{\ALICE}{$r_3$} - t\BY{\ALICE}{$r_2$} }{ t\BY{\ALICE}{$e_3$} - t\BY{\ALICE}{$e_2$} }=4,
\end{equation}
and thus determine $k=2$.
So, there must \textit{two} of Bob's clock-ticks between events $B_2$ and $B_3$ on his worldline.
Since the sides of the 
light-clock diamonds
are traced out by light-signals 
(drawn parallel to the lines of the rotated grid), 
one is led to drawing \textit{two} congruent 
causal diamonds, each a prototype of Bob's light-clock diamond.

We then ask the student to somehow characterize Bob's 
light-clock diamond
so as to avoid having to
always do a radar-experiment to construct it.
By being able to count squares on the rotated grid, 
it is realized that Bob's
light-clock diamond
has the same area as Alice's.
Indeed, the edges of 
Bob's 
light-clock diamond
have sizes $k$ and $1/k$, 
compared to the corresponding edges of Alice's 
light-clock diamond.
We can now draw Bob's 
light-clock diamonds
along his entire worldline.

Furthermore, the student is led to notice that the 
causal diamond 
from $B_2$ to $B_3$ 
(two events on Bob's worldline) has the same shape as one of Bob's
light-clock diamonds,
and 
that the area of that causal diamond 
is equal to the square of the proper-time interval from $B_2$ to $B_3$.
Note that both Alice and Bob, each using their 
light-clock diamonds
can partition the 
area of that causal diamond into $k^2=4$ 
light-clock diamond areas,
which also displays the invariance of the square-interval.

These conclusions can be reinforced by repeating the construction for 
another numerically convenient velocity, 
say, $\VEL=(4/5)\cspeed$.

\subsection{Bob's Coordinate Systems}
\label{sec:BobCoordinates}

With the prototype for Bob's light-clock diamonds, 
it is now easy to construct Bob's rectangular and light-cone coordinate systems,
in complete analogy to Alice's coordinate systems.
(Refer to Figures ~\ref{fig:BobCoords} and~\ref{fig:BobLightConeCoords} and their captions,
and compare with Figures~\ref{fig:AliceCoords} and~\ref{fig:AliceLightConeCoords}.)

As mentioned in Sec.~\ref{sec:proofForEducators}, 
three relativistic effects are evident from the diagram.
We will discuss the \textit{time-dilation} 
and \textit{length-contraction} effects
in the Applications 
(Sections~\ref{sec:TimeDilation}, \ref{sec:Symmetry}, and~\ref{sec:LengthContraction}).
Here, we address the remaining effect.

A striking feature of Bob's rectangular coordinate system (in Fig.~\ref{fig:BobCoords})
is that Bob's $x$-axis is not parallel with Alice's $x$-axis.
Since an observer's $x$-axis represents a set of events that 
are simultaneous with the event at $t=0$ for that observer,
the spacetime diagram indicates that these observers will disagree on
whether two distinct events (say, events $O$ and $G$) are simultaneous.
This is the \textit{relativity of simultaneity}.

A related feature is that Bob's $x$-axis is not perpendicular---in the familiar
Euclidean sense---to Bob's $t$-axis, as it appears to be for Alice's $x$- and $t$-axes.
\cite{fn:perpendicular}
This feature is an indication that the geometry of this spacetime diagram
is not a Euclidean geometry, but a Minkowskian geometry.%
\cite{Minkowski}
One important difference is that the notion of perpendicularity in a spacetime diagram
is not specified by a circle, but by a causal diamond (which encodes
some properties of the Minkowskian analogue of a circle, the hyperbola).
From the discussion in Sec.~\ref{sec:UsefulTerminology},
since Bob's $x$- and $t$-axes are parallel to the 
diagonals of some causal diamond (specifically, Bob's light-clock diamond), 
those axes are in fact
``spacetime-perpendicular.''

We will apply our construction to some examples from Special Relativity 
(in Sec.~\ref{sec:Applications})
 to help develop skills for
reasoning physically and geometrically with spacetime diagrams.

\begin{figure}[!Hh]

\def\aliceDTSCALE{\psscalebox{1 1}}
\def\aliceDXSCALE{\psscalebox{1 1}}
\def\aliceDT{\multirput[bl]{0}(0,0)(10,10){7}}%
\def\aliceDX{\multirput[ur]{-90}(0,0)(10,-10){6}}%
\def\aliceDTHATCH{\aliceDT{\aliceDTSCALE{\pstickHATCH{\AliceColor!50}}}}%
\def\aliceDTLINE{\aliceDT{\aliceDTSCALE{\pstickLINE{\AliceColor}}}}%
\def\aliceDTCONETHICK{\aliceDT{\aliceDTSCALE{\pstickCONETHICK{\AliceColor}}}}%
\def\aliceDXHATCH{\aliceDX{\aliceDXSCALE{\pstickHATCH{\AliceColor!50}}}}%
\def\aliceDXLINE{\aliceDX{\aliceDXSCALE{\pstickLINE{\AliceColor!75}}}}%
\def\aliceDXCONE{\aliceDX{\aliceDXSCALE{\pstickCONE{\AliceColor!75}}}}%

\def\bobDTSCALE{\psscalebox{2 .5}}
\def\bobDXSCALE{\psscalebox{.5 2}}
\def\bobDT{\multirput[bl]{0}(0,0)(20,5){7}}%
\def\bobDX{\multirput[ur]{-90}(0,0)(20,-5){6}}%
\def\bobDTHATCH{\bobDT{\bobDTSCALE{\pstickHATCH{\BobColor}}}}%
\def\bobDTLINE{\bobDT{\bobDTSCALE{\pstickLINE{\BobColor}}}}%
\def\bobDTCONE{\bobDT{\bobDTSCALE{\pstickCONE{\BobColor}}}}%
\def\bobDXHATCH{\bobDX{\bobDXSCALE{\pstickHATCH{\BobColor!75}}}}%
\def\bobDXLINE{\bobDX{\bobDXSCALE{\pstickLINE{\BobColor!75}}}}%
\def\bobDXCONE{\bobDX{\bobDXSCALE{\pstickCONE{\BobColor!75}}}}%

{%
\psset{unit=.85}
\begin{pspicture}(-50,0)(50,112)%
\psclip{\psframe[linestyle=none](-50,-10.25)(50,92)}{%
\rput[bl]{45}(0,0){%
\multirput[bl]{0}(0,-50)(0,10){20}{%
\multirput[bl]{0}(-50,0)(10,0){20}{\pstickSQUARE{black!75}}}}}%
\endpsclip%
\psclip{\psframe[linestyle=none](-50,-12)(50,92)}
{%
\rput[bl]{45}(-14.142,0)
{%
\aliceDTHATCH%
\aliceDXHATCH%
\aliceDTLINE%
\aliceDXLINE%
\bobDTHATCH%
\bobDXHATCH%
\bobDTLINE%
\bobDXLINE%
\psline[linecolor=white!25,linewidth=.3]{c-c}(20,0)(0,0)
\pscircle(80,20){.75}\rput[l]{-45}(80,20){\ $Q$}%
\rput[b]{-45}(65,30){\rput[l]{55.0}(0,0){\psframebox[linecolor=black!50,framearc=.3]{\psframebox*[framearc=.3,fillcolor=white,opacity=.75]{\begin{tabular}{c}Bob's\\$t$-axis\end{tabular}}}}}%
\rput[b]{-45}(35,-25){\rput[l]{30.9}(-8,10){\psframebox[linecolor=black!50,framearc=.3]{\psframebox*[framearc=.3,fillcolor=white,opacity=.75]{\begin{tabular}{c}Bob's\\$x$-axis\end{tabular}}}}}%
\pscircle(20,5){.75}\rput[c]{-45}(20,5){\rput[c](-6,-3){\multirput[b]{0}(-0.3,.2)(.1,-.1){6}{\color{white}$F$}\rput[b](0,0){$F$}}}%
\pscircle(20,-5){.75}\rput[t]{-45}(20,-5){\rput[lt](-1,-9){\multirput[b]{0}(-0.3,.2)(.1,-.1){6}{\color{white}$G$}\rput[b](0,0){$G$}}}%
\pscircle(0,5){.75}\rput[t]{-45}(-2,7){$Y\ $}%
\pscircle(20,0){.75}\rput[c]{-45}(20,0){\rput[c](4,0){\multirput[b]{0}(-0.3,.2)(.1,-.1){6}{\color{white}$Z$}\rput[b](0,0){$Z$}}}%
\pscircle(0,0){.75}\rput[t]{-45}(0,0){\rput[t](-.5,-2){$O$}}%
}%
}%
\endpsclip%
\end{pspicture}%
}
\caption{
Bob's Rectangular Coordinate System uses the {\it diagonals} 
of his light-clock diamonds to locate events. 
Bob's time-axis is marked by a string of light-clock diamonds 
with their timelike-diagonals along $\VEC{OF}$ on his worldline.
Bob's spatial $x$-axis is marked by a string of light-clock diamonds 
with their spacelike-diagonals along $\VEC{OG}$,
on a line of simultaneity 
according to Bob (parallel to $YZ$)
through event $O$.
From the diagram, since $\VEC{OQ}=4\,\VEC{OF}+0\,\VEC{OG}$,
Bob assigns event~$Q$ rectangular coordinates 
$(t\BY{\BOB}{Q},x\BY{\BOB}{Q})=(4,0)$.}
\label{fig:BobCoords}
\end{figure}
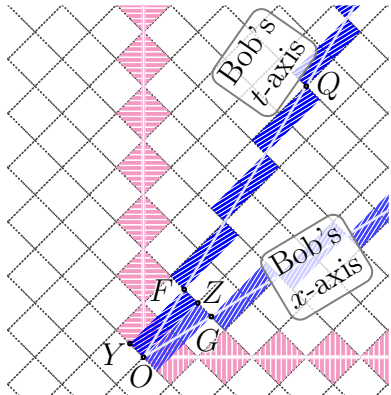

\begin{figure}[!Hh]                                              

\def\bobDTSCALE{\psscalebox{2 .5}}
\def\bobDT{\multirput[bl]{0}(0,0)(20,5){1}}%
\def\bobDTHATCH{\bobDT{\bobDTSCALE{\pstickHATCH{\BobColor!50}}}}%
\def\bobDTCONE{\bobDT{\bobDTSCALE{\pstickCONE{\BobColor}}}}%
\def\bobDTCONETHICK{\bobDT{\bobDTSCALE{\pstickCONETHICK{\BobColor}}}}%

\def\bobDVSCALE{\psscalebox{2 .5}}
\def\bobDUSCALE{\psscalebox{.5 2}}
\def\bobDV{\multirput[bl]{0}(-20,0)(0,5){11}}%
\def\bobDU{\multirput[br]{-90}(0,0)(20,0){5}}%
\def\bobDVHATCH{\bobDV{\bobDVSCALE{\pstickHATCHLV{\BobColor!75}}\psframe[linecolor=white,linestyle=solid,dotsep=.2929pt,linewidth=.3071pt](0,0)(20,5)}}%
\def\bobDVLINE{\bobDV{\bobDVSCALE{\pstickLINELV{\BobColor}}}}%
\def\bobDVCONETHICK{\bobDV{\aliceDVSCALE{\pstickCONETHICK{\BobColor}}}}%
\def\bobDUHATCH{\bobDU{\bobDUSCALE{\pstickHATCHLU{\BobColor!75}}\psframe[linecolor=white,linestyle=solid,dotsep=.2929pt,linewidth=.3071pt](0,0)(5,20)}}%
\def\bobDULINE{\bobDU{\bobDUSCALE{\pstickLINELU{\BobColor!75}}}}%
\def\bobDUCONE{\bobDU{\bobDUSCALE{\pstickCONE{\BobColor!75}}}}%

{%
\psset{unit=.85}
\begin{pspicture*}(-50,-14.14)(50,92)%
\psclip{\psframe[linestyle=none](-50,-12.14)(50,92)}{%
\rput[bl]{45}(0,0){%
\multirput[bl]{0}(0,-50)(0,10){20}{%
\multirput[bl]{0}(-50,0)(10,0){20}{\pstickSQUARE{black!75}}}}}%
\endpsclip%
\psclip{\psframe[linestyle=none](-50,-12)(50,92)}
{%
\rput[bl]{45}(-14.142,0)
{%
\bobDTHATCH%
\bobDTLINE%
\bobDTCONETHICK%
\rput(0.,0.){
\psline[linecolor=\BobColor,linewidth=.6pt](0,1)(100,1)
\psline[linecolor=\BobColor,linewidth=.6pt](1,0)(1,100)
\psline[linecolor=white,linewidth=5pt,doubleline=false,doublesep=2pt](-1,-1)(100,-1)
\psline[linecolor=white,linewidth=5pt,doubleline=false,doublesep=2pt](-1,-1)(-1,100)
}
\bobDVHATCH%
\bobDUHATCH%
\pscircle(80,20){.75}\rput[l]{-45}(80,20){\ $Q$}%
\rput[b]{-45}(0,50){\rput[l]{45}(5.5,-3){\psframebox[linecolor=black!50,framearc=.3]{\psframebox*[framearc=.3,fillcolor=white,opacity=.75]{\begin{tabular}{c}Bob's\\$v$-axis\end{tabular}}}}}%
\rput[b]{-45}(30,-25){\rput[l]{45}(0,5){\psframebox[linecolor=black!50,framearc=.3]{\psframebox*[framearc=.3,fillcolor=white,opacity=.75]{\begin{tabular}{c}Bob's\\$u$-axis\end{tabular}}}}}%
\pscircle[linecolor=white](20,5){1.5}\pscircle(20,5){.75}\rput[c]{-45}(20,5){\rput[c](-4,-3){\multirput[b]{0}(-0.3,.2)(.1,-.1){6}{\color{white}$F$}\rput[b](0,0){$F$}}}%
\pscircle[linecolor=white](20,-5){1.5}\pscircle(20,-5){.75}\rput[t]{-45}(20,-5){\rput[lt](0,-7){\multirput[b]{0}(-0.3,.2)(.1,-.1){6}{\color{white}$G$}\rput[b](0,0){$G$}}}%
\pscircle[linecolor=white](0,5){1.5}\pscircle(0,5){.75}\rput[b]{-45}(-1,4){\rput[l](1,5){\multirput[b]{0}(-0.3,.2)(.1,-.1){6}{\color{white}$Y$}\rput[b](0,0){$Y$}}}%
\pscircle[linecolor=white](20,0){1.5}\pscircle(20,0){.75}\rput[b]{-45}(20,0){\rput[c](1,5){\multirput[b]{0}(-0.3,.2)(.1,-.1){6}{\color{white}$Z$}\rput[b](0,0){$Z$}}}%
\pscircle[linecolor=white](0,0){1.5}\pscircle(0,0){.75}\rput[t]{-45}(0,0){\rput[t](-.5,-3){$O$}}%

}%
}%
\endpsclip%
\end{pspicture*}%
}
\caption{
Bob's Light-Cone Coordinate System uses the {\it edges} of his light-clock diamonds to locate events. 
Since 
$\VEC{OF}=\VEC{OZ}+\VEC{OY}$ and $\VEC{OG}=\VEC{OZ}-\VEC{OY}$,
we have $\VEC{OQ}=4\,\VEC{OZ}+4\,\VEC{OY}$.
So, Bob assigns
event $Q$ light-cone coordinates $(u\BY{\BOB}{Q},v\BY{\BOB}{Q})=(4,4)$.
Analogous to the situation with Alice, Bob's $u$- and $v$-axes are marked 
by strings of light-clock diamonds  with their lightlike-edges along 
$\VEC{OZ}$ and $\VEC{OY}$, respectively, on the light-cone of event $O$.}
\label{fig:BobLightConeCoords}
\end{figure}
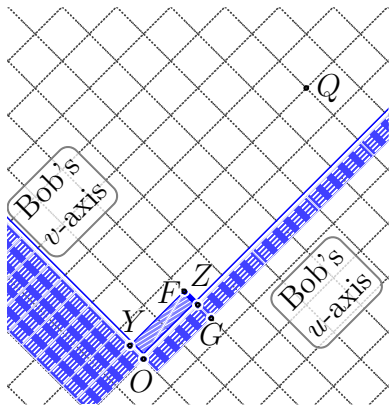

\subsection{Subdivided grids}
\label{sec:subdivided}

As we have seen,
when the relative-velocity $\VEL$ of an inertial observer is rational, 
say $\VEL=\Delta x/\Delta t$ where $\Delta x$ and $\Delta t$ are integers, the worldline of that observer
can be constructed by using a Minkowski-right triangle formed by 
counting off $\Delta t$ of Alice's temporally-arranged diamonds,
followed by 
counting off $\Delta x$ of Alice's spatially-arranged diamonds.
Since the square-interval of the hypotenuse is the integer $\Delta s^2=(\Delta t)^2-(\Delta x)^2$,
the area of its causal diamond can be determined by counting grid squares.
 
To go further and count off an integer number $\Delta s$ 
of diamonds along the hypotenuse,
the square-interval should be a perfect square.
This occurs when the Bondi $k$-factor $k=\sqrt{ (1+\VEL)/(1-\VEL) }$
(Eq.~(\ref{eq:BondiKfactor})%
) is rational, with value $\Delta s/(\Delta t-\Delta x)$.
(These restrictions can be formulated in terms
of Pythagorean triples.%
\cite{fn:PythagoreanTriples})

When the $k$-factors are ratios of small integers,
it becomes easy to graphically construct the 
diamonds 
of the observers, 
especially if the grid is suitably subdivided.
A $6\times 6$ subdivision is useful since it can accomodate
the $k$-factors $1$, $2$, $3/2$, and $3$ and their reciprocals,
which correspond to the velocities $0$, $\pm 3/5$, $\pm 5/13$, and $\pm 4/5$.

To handle arbitrary relative-velocities, 
one can advance to the use of algebraic formulas [derived later],
whose meanings would have been motivated by the special cases noted above.

\section{Applications}
\label{sec:Applications}

We now offer a series of standard calculations in special relativity
with our graphical method. These are concise summaries
of what has been presented to students through a series of worksheets.
Although we proceed in a sequence to highlight ``relativistic effects,''
we regard the operational radar-measurement methods in Sec.~\ref{sec:Radar} 
to be a more fundamental starting point.

\subsection{Time dilation}
\label{sec:TimeDilation}

{\it 
After leaving inertial observer Alice at event $O$,
another inertial observer Bob travels at $(3/5)\cspeed$ according to Alice.
Thus, according to Alice, after 5 of her ticks have elapsed, 
Bob is located 3 of her ticks away, at event $Q$.
How much of Bob's proper time has elapsed from events $O$ to $Q$, both on his worldline?}

\def\aliceDTSCALE{\psscalebox{1 1}}
\def\aliceDXSCALE{\psscalebox{1 1}}
\def\aliceDT{\multirput[bl]{0}(0,0)(10,10){5}}%
\def\aliceDX{\multirput[ur]{-90}(50,50)(10,-10){3}}%
\def\aliceDTHATCH{\aliceDT{\aliceDTSCALE{\pstickHATCH{\AliceColor}}}}%
\def\aliceDTLINE{\aliceDT{\aliceDTSCALE{\pstickLINE{\AliceColor}}}}%
\def\aliceDTCONE{\aliceDT{\aliceDTSCALE{\pstickCONE{\AliceColor}}}}%
\def\aliceDXHATCH{\aliceDX{\aliceDXSCALE{\pstickHATCH{\AliceColor!75}}}}%
\def\aliceDXLINE{\aliceDX{\aliceDXSCALE{\pstickLINE{\AliceColor!75}}}}%
\def\aliceDXCONE{\aliceDX{\aliceDXSCALE{\pstickCONE{\AliceColor!75}}}}%

\def\bobDTSCALE{\psscalebox{2 .5}}
\def\bobDXSCALE{\psscalebox{.5 2}}
\def\bobDT{\multirput[bl]{0}(0,0)(20,5){4}}%
\def\bobDTHATCH{\bobDT{\bobDTSCALE{\pstickHATCH{\BobColor}}}}%
\def\bobDTLINE{\bobDT{\bobDTSCALE{\pstickLINE{\BobColor}}}}%
\def\bobDTCONE{\bobDT{\bobDTSCALE{\pstickCONE{\BobColor}}}}%

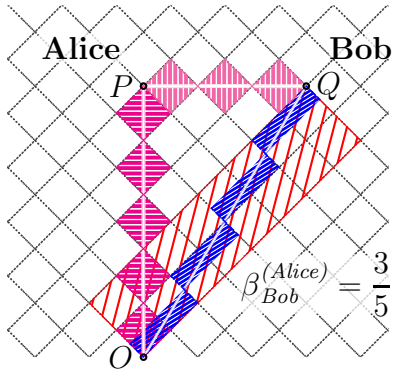
\begin{figure}[!Hh]
{%
\psset{unit=.85}
\begin{pspicture}(-50,0)(50,92)%
\psclip{\psframe[linestyle=none](-50,-0.25)(50,92)}{%
\rput[bl]{45}(0,0){%
\multirput[bl]{0}(0,-50)(0,10){20}{%
\multirput[bl]{0}(-50,0)(10,0){20}{\pstickSQUARE{black!75}}}}}%
\endpsclip%
\rput[bl]{45}(-14.142,0)
{%
\rput[b]{-45}(50,-11){\rput[t](2,-1){\psframebox*[boxsep=true,framesep=-1pt,framearc=.3,fillcolor=white,opacity=.75]{\begin{tabular}{c}$\VEL\BY{\ALICE}{\Bob}=\displaystyle\frac{3}{5}\cspeed$\end{tabular}}}}%
\rput[bl]{0}(0,0){\pstickDIAMOND{red}{80,20}}%
\aliceDTHATCH%
\aliceDXHATCH%
\bobDTHATCH%
\aliceDTLINE%
\aliceDXLINE%
\bobDTLINE%
\aliceDTCONE%
\aliceDXCONE%
\bobDTCONE%
\pscircle(0,0){1}\rput[r]{-45}(0,0){$O\ $}%
\pscircle(50,50){1}\rput[r]{-45}(50,50){$P\ $}%
\pscircle(80,20){1}\rput[l]{-45}(80,20){$\ Q$}%
\rput[b]{-45}(50,50){\rput[r](-5,10){\psframebox*[framearc=.3,fillcolor=white,opacity=.75,framesep=1]{\bf Alice}}}%
\rput[b]{-45}(80,20){\rput[l](6,10){\psframebox*[framearc=.3,fillcolor=white,opacity=.75]{\bf Bob}}}%
}%
\end{pspicture}%
}
\caption{\label{fig:TimeDilation} Time Dilation: $t\BY\ALICE{Q}=t\BY\ALICE{P} > t\BY\BOB{Q}$.
}
\end{figure}

In Fig.~\ref{fig:TimeDilation}, 
we use Alice's diamonds to draw Bob's worldline, 
with $OP$ and $PQ$ chosen so that $(PQ)/(OP)=\beta=3/5$.
Since the causal diamond of $OQ$ has $OQ$ as the timelike diagonal,
the proper-time elapsed along $OQ$ is equal to the square root of the area of that causal diamond.
From the diagram, we count the area of the causal diamond to be $16$, 
so that the proper-time is $\sqrt{16}=4$ ticks.
We can now draw 4 of Bob's light-clock diamonds.
Note that while Bob declares the elapsed time between
 the events $O$ and $Q$ that he experiences is $4$ ticks,
Alice declares the elapsed time to be $5$ ticks. 
This is the time dilation effect, with time-dilation factor $\gamma=(OP)/(OQ)=5/4$.
This agrees with the formula $\gamma=(1-\overc{\VEL}^2)^{-1/2}=5/4$ for $\VEL=(3/5)\cspeed$,
which we explicitly use in Sec.~\ref{sec:BoostTransformation}, and 
derive in the Appendix.

More generally, in the timelike-future of $O$, 
consider two events, $P$ and $Q$, that Alice determines to be simultaneous.
The ratio of the areas of the causal diamonds of $OP$ (not shown) and of $OQ$ is equal to 
$\gamma^2$, the square of the time-dilation factor.

\def\aliceDTSCALE{\psscalebox{1 1}}
\def\aliceDXSCALE{\psscalebox{1 1}}
\def\aliceDT{\multirput[bl]{0}(0,0)(10,10){5}}%
\def\aliceDX{\multirput[ur]{-90}(50,50)(10,-10){3}}%
\def\aliceDTHATCH{\aliceDT{\aliceDTSCALE{\pstickHATCH{\AliceColor}}}}%
\def\aliceDTLINE{\aliceDT{\aliceDTSCALE{\pstickLINE{\AliceColor}}}}%
\def\aliceDTCONE{\aliceDT{\aliceDTSCALE{\pstickCONE{\AliceColor}}}}%
\def\aliceDXHATCH{\aliceDX{\aliceDXSCALE{\pstickHATCH{\AliceColor!75}}}}%
\def\aliceDXLINE{\aliceDX{\aliceDXSCALE{\pstickLINE{\AliceColor!75}}}}%
\def\aliceDXCONE{\aliceDX{\aliceDXSCALE{\pstickCONE{\AliceColor!75}}}}%

\def\bobDTSCALE{\psscalebox{2 .5}}
\def\bobDXSCALE{\psscalebox{.5 2}}
\def\bobDT{\multirput[bl]{0}(0,0)(20,5){5}}%
\def\bobDX{\multirput[ur]{90}(100,25)(-20,5){3}}%
\def\bobDTHATCH{\bobDT{\bobDTSCALE{\pstickHATCH{\BobColor}}}}%
\def\bobDTLINE{\bobDT{\bobDTSCALE{\pstickLINE{\BobColor}}}}%
\def\bobDTCONE{\bobDT{\bobDTSCALE{\pstickCONE{\BobColor}}}}%
\def\bobDXHATCH{\bobDX{\bobDXSCALE{\pstickHATCH{\BobColor!75}}}}%
\def\bobDXLINE{\bobDX{\bobDXSCALE{\pstickLINE{\BobColor!75}}}}%
\def\bobDXCONE{\bobDX{\bobDXSCALE{\pstickCONE{\BobColor!75}}}}%

\subsection{Symmetry of the Inertial Observers}
\label{sec:Symmetry}

Refer to Figure~\ref{fig:Symmetry}.
In accordance with the Relativity Principle,
for two events, $O$ and $Q'$, experienced by Alice, 
Bob will observe a longer time-interval between those events 
than Alice will---with the same time-dilation factor $\gamma=5/4$.
Using Bob's diamonds, we construct the analogue of Alice's diagram.
Since the small diagram we have been using becomes 
a little cluttered when displaying the symmetry, 
we have included additional pairs of events $\{H,J\}$ and $\{H',J'\}$ in Fig.~\ref{fig:Symmetry}.

\def\aliceDTSCALE{\psscalebox{1 1}}
\def\aliceDXSCALE{\psscalebox{1 1}}
\def\aliceDTB{\multirput[bl]{0}(0,0)(10,10){10}}%
\def\aliceDXB{\multirput[ur]{-90}(100,100)(10,-10){6}}%
\def\aliceDTBHATCH{\aliceDTB{\aliceDTSCALE{\pstickHATCH{\AliceColor}}}}%
\def\aliceDTBLINE{\aliceDTB{\aliceDTSCALE{\pstickLINE{\AliceColor}}}}%
\def\aliceDTBCONE{\aliceDTB{\aliceDTSCALE{\pstickCONE{\AliceColor}}}}%
\def\aliceDXBHATCH{\aliceDXB{\aliceDXSCALE{\pstickHATCH{\AliceColor!75}}}}%
\def\aliceDXBLINE{\aliceDXB{\aliceDXSCALE{\pstickLINE{\AliceColor!75}}}}%
\def\aliceDXBCONE{\aliceDXB{\aliceDXSCALE{\pstickCONE{\AliceColor!75}}}}%

\def\bobDTSCALE{\psscalebox{2 .5}}
\def\bobDXSCALE{\psscalebox{.5 2}}
\def\bobDTB{\multirput[bl]{0}(0,0)(20,5){10}}%
\def\bobDXB{\multirput[ur]{90}(200,50)(-20,5){6}}%
\def\bobDTBHATCH{\bobDTB{\bobDTSCALE{\pstickHATCH{\BobColor}}}}%
\def\bobDTBLINE{\bobDTB{\bobDTSCALE{\pstickLINE{\BobColor}}}}%
\def\bobDTBCONE{\bobDTB{\bobDTSCALE{\pstickCONE{\BobColor}}}}%
\def\bobDXBHATCH{\bobDXB{\bobDXSCALE{\pstickHATCH{\BobColor!75}}}}%
\def\bobDXBLINE{\bobDXB{\bobDXSCALE{\pstickLINE{\BobColor!75}}}}%
\def\bobDXBCONE{\bobDXB{\bobDXSCALE{\pstickCONE{\BobColor!75}}}}%

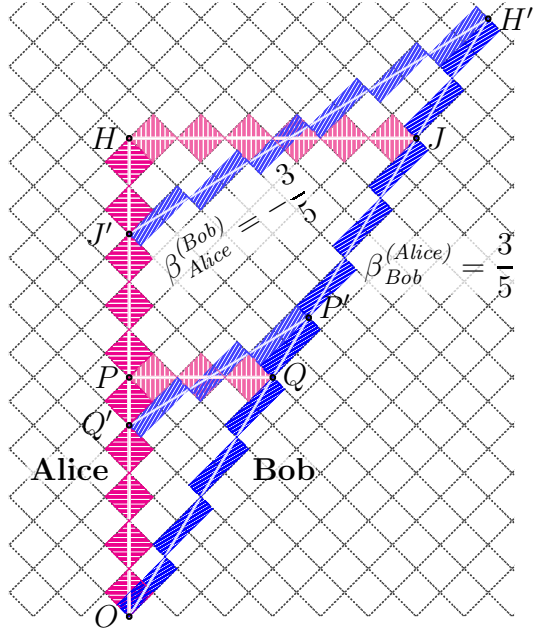
\begin{figure}[!Hht]
{%
\psset{unit=.75}
\begin{pspicture}(-50,0)(100,182)%
\psclip{\psframe[linestyle=none](-50,-0.25)(100,182)}{%
\rput[bl]{45}(0,0){%
\multirput[bl]{0}(0,-80)(0,10){25}{%
\multirput[bl]{0}(-50,0)(10,0){25}{\pstickSQUARE{black!75}}}}}%
\endpsclip%
\rput[bl]{45}(-14.142,0)
{%
\aliceDTHATCH%
\aliceDXHATCH%
\aliceDTBHATCH%
\aliceDXBHATCH%
\bobDTHATCH%
\bobDXHATCH%
\bobDTBHATCH%
\bobDXBHATCH%
\aliceDTLINE%
\aliceDXLINE%
\aliceDTBLINE%
\aliceDXBLINE%
\bobDTLINE%
\bobDXLINE%
\bobDTBLINE%
\bobDXBLINE%
\aliceDTCONE%
\aliceDXCONE%
\aliceDTBCONE%
\aliceDXBCONE%
\bobDTCONE%
\bobDXCONE%
\bobDTBCONE%
\bobDXBCONE%
\pscircle(50,50){1}\rput[r]{-45}(50,50){$P\ $}
\pscircle(80,20){1}\rput[l]{-45}(80,20){$\ Q$}
\pscircle(100,25){1}\rput[bl]{-45}(100,25){$\ P'$ }
\pscircle(40,40){1}\rput[r]{-45}(40,40){$Q'\,$ }
\pscircle(0,0){1}\rput[r]{-45}(0,0){$O\ $}%
\pscircle(100,100){1}\rput[r]{-45}(100,100){$H\ $}%
\pscircle(160,40){1}\rput[l]{-45}(160,40){$\ J$}%
\pscircle(200,50){1}\rput[l]{-45}(200,50){$\ H'$ }%
\pscircle(80,80){1}\rput[r]{-45}(80,80){$J'\,$ }%
\rput[b]{-45}(160,40){\rput[t](7,-29){\psframebox*[boxsep=true,framesep=-1pt,framearc=.3,fillcolor=white,opacity=.75]{\begin{tabular}{c}$\VEL\BY{\ALICE}{\Bob}=\displaystyle\frac{3}{5}\cspeed$\end{tabular}}}}%
\rput[b]{-45}(100,55){\rput[t]{30.9}(-5,12){\psframebox*[boxsep=true,framesep=-1pt,framearc=.3,fillcolor=white,opacity=.75]{\begin{tabular}{c}$\VEL\BY{\BOB}{\Alice}=\displaystyle-\frac{3}{5}\cspeed$\end{tabular}}}}%
\rput[b]{-45}(50,50){\rput[r](-5,-27){\psframebox*[framearc=.3,fillcolor=white,opacity=.75,framesep=1]{\bf Alice}}}%
\rput[b]{-45}(80,20){\rput[l](-6,-27){\psframebox*[framearc=.3,fillcolor=white,opacity=.75]{\bf Bob}}}%
}%
\end{pspicture}%
}
\caption{\label{fig:Symmetry} Symmetry of time-dilation.
}
\end{figure}

Note that the use of similarity in this diagram suggests that
problems involving this relative-speed
with \textit{arbitrary} lengths and time can be handled by setting up
proportions.

\subsection{The Clock Effect}

\def\aliceDTSCALE{\psscalebox{1 1}}
\def\aliceDXSCALE{\psscalebox{1 1}}
\def\aliceDT{\multirput[bl]{0}(0,0)(10,10){10}}%
\def\aliceDX{\multirput[ur]{-90}(50,50)(10,-10){3}}%
\def\aliceDTHATCH{\aliceDT{\aliceDTSCALE{\pstickHATCH{\AliceColor}}}}%
\def\aliceDTLINE{\aliceDT{\aliceDTSCALE{\pstickLINE{\AliceColor}}}}%
\def\aliceDTCONE{\aliceDT{\aliceDTSCALE{\pstickCONE{\AliceColor}}}}%
\def\aliceDXHATCH{\aliceDX{\aliceDXSCALE{\pstickHATCH{\AliceColor!75}}}}%
\def\aliceDXLINE{\aliceDX{\aliceDXSCALE{\pstickLINE{\AliceColor!75}}}}%
\def\aliceDXCONE{\aliceDX{\aliceDXSCALE{\pstickCONE{\AliceColor!75}}}}%

\def\bobDTSCALE{\psscalebox{2 .5}}
\def\bobDXSCALE{\psscalebox{.5 2}}
\def\bobDTBSCALE{\psscalebox{-.5 -2}}
\def\bobDT{\multirput[bl]{0}(0,0)(20,5){4}}%
\def\bobDTB{\multirput[bl]{180}(80,20)(5,20){4}}%
\def\bobDTHATCH{\bobDT{\bobDTSCALE{\pstickHATCH{\BobColor}}}}%
\def\bobDTLINE{\bobDT{\bobDTSCALE{\pstickLINE{\BobColor}}}}%
\def\bobDTCONE{\bobDT{\bobDTSCALE{\pstickCONE{\BobColor}}}}%

\def\bobDTBHATCH{\bobDTB{\bobDTBSCALE{\pstickHATCH{\BobColor}}}}%
\def\bobDTBLINE{\bobDTB{\bobDTBSCALE{\pstickLINE{\BobColor}}}}%
\def\bobDTBCONE{\bobDTB{\bobDTBSCALE{\pstickCONE{\BobColor}}}}%

{\it
After leaving Alice at event $O$,
Bob travels away at velocity $(3/5)\cspeed$, 
instantaneously turning around at a distant event $Q$ 
and returning with velocity $-(3/5)\cspeed$.
At the reunion event $Z$, 
if Alice had aged $10$ ticks since the separation event,
how much did Bob age?}

\begin{figure}[!Hh]
{%
\psset{unit=.85}
\begin{pspicture}(-50,0)(50,152)%
\psclip{\psframe[linestyle=none](-50,-0.25)(50,142)}{%
\rput[bl]{45}(0,0){%
\multirput[bl]{0}(0,-50)(0,10){20}{%
\multirput[bl]{0}(-50,0)(10,0){20}{\pstickSQUARE{black!75}}}}}%
\endpsclip%
\rput[bl]{45}(-14.142,0)
{%
\rput[bl]{0}(0,0){\pstickDIAMOND{red}{80,20}}%
\aliceDTHATCH%
\aliceDXHATCH%
\bobDTHATCH%
\bobDTBHATCH%
\aliceDTLINE%
\aliceDXLINE%
\bobDTLINE%
\bobDTBLINE%
\aliceDTCONE%
\aliceDXCONE%
\bobDTCONE%
\bobDTBCONE%
\pscircle(0,0){1}\rput[r]{-45}(0,0){$O\ $}%
\pscircle(50,50){1}\rput[r]{-45}(50,50){$P\ $}%
\pscircle(80,20){1}\rput[l]{-45}(80,20){$\ Q$}%
\pscircle(100,100){1}\rput[r]{-45}(100,100){$Z\ $}%
\rput[b]{-45}(55,-10){\rput[r](20,-10){\psframebox*[boxsep=true,framesep=-1pt,framearc=.3,fillcolor=white,opacity=.75]{\begin{tabular}{c}$\VEL\BY{\ALICE}{\Bob,out}=\displaystyle\frac{3}{5}\cspeed$\end{tabular}}}}%
\rput[b]{-45}(55,-10){\rput[r](20,91){\psframebox*[boxsep=true,framesep=-1pt,framearc=.3,fillcolor=white,opacity=.75]{\begin{tabular}{c}$\VEL\BY{\ALICE}{\Bob,in}=\displaystyle-\frac{3}{5}\cspeed$\end{tabular}}}}%
\rput[b]{-45}(50,50){\rput[r](-5,15){\psframebox*[framearc=.3,fillcolor=white,opacity=.75,framesep=1]{\bf Alice}}}%
\rput[b]{-45}(80,20){\rput[l](6,15){\psframebox*[framearc=.3,fillcolor=white,opacity=.75]{\bf Bob}}}%
}%
\end{pspicture}%
}
\caption{\label{fig:ClockEffect} Clock Effect. 
Alice ages $10\rm\ ticks$ along her inertial worldline from $O$ to $Z$, 
whereas Bob (traveling to and from event $Q$ with speed $(3/5)\cspeed$)
ages $4+4=8\rm\ ticks$ along his non-inertial worldline from $O$ to $Q$ to $Z$.
}
\end{figure}
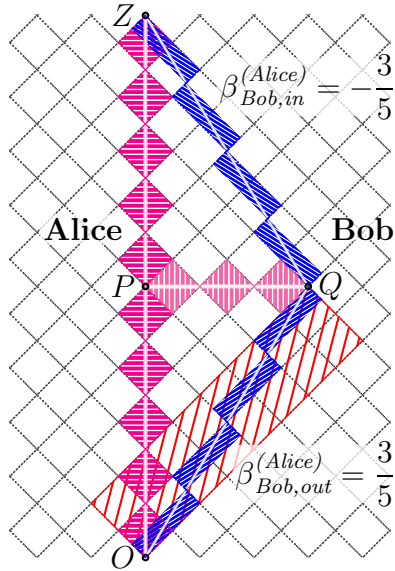

Refer to Fig.~\ref{fig:ClockEffect}.
Determine Bob's diamonds by treating 
his non-inertial trip as piecewise-inertial segments.
By constructing the causal diamonds from $O$ to $Q$,
then from $Q$ to $Z$, we determine from the diagram
$\sqrt{16}$ ticks along $OQ$,
followed by another $\sqrt{16}$ diamonds along $QZ$, 
for a total of $8$ ticks.
This does not equal the $10$ ticks logged by Alice's [inertial] trip from $O$ to $Z$.
This route-dependence of elapsed proper-time between two events is the Clock Effect.

\subsection{Length Contraction}
\label{sec:LengthContraction}

{\it
Bob traveling at $(3/5)\cspeed$ carries a ladder that is $5\rm\  ticks$ long.
How long is that ladder according to Alice?
}

Refer to Fig.~\ref{fig:LengthContraction}.
We use Alice's diamonds to draw Bob's worldline and 
his diamonds, 
with $OP$ and $PQ$ chosen so that $(PQ)/(OP)=3/5$.
Then, use Bob's light-clock diamonds to measure from event $O$ to event $Y$, 
a spatial displacement of $5\rm\ ticks$, the ladder's rest-length.
(Note that Bob regards events $O$ and $Y$ as simultaneous.)
Through event $Y$, construct the worldline of the ladder's far end 
as a parallel to Bob's worldline.

Alice determines the length of Bob's ladder by finding the spatial distance between
events $O$ and $X$, where $X$ is the event on the worldline of the ladder's far end 
that Alice regards as simultaneous with $O$. Since $OX$ is the spatial-diagonal
of the causal diamond between $O$ and $X$, the length of the ladder according to Alice
is square-root of the area of the causal diamond between $O$ and $X$: $(OX)=4\rm\ ticks$.
Since $(OX) < (OY)$, the observed-length of the ladder is shorter than its rest-length.
This is the length-contraction effect, with 
length-contraction factor equal to the time-dilation factor $\gamma=5/4$.

By symmetry,
Bob determines the length of Alice's identical ladder, 
with its far end along a worldline (not shown) through events $X'$ and $Y'$,
to be shorter than its rest-length by the same length-contraction factor.
[A triangle like $OX'Y'$ will be used in a proof 
for the Lorentz Boost transformation formula shown in 
Fig.~\ref{fig:BoostTransformationDerivation}.]

\def\aliceDTSCALE{\psscalebox{1 1}}
\def\aliceDXSCALE{\psscalebox{1 1}}
\def\aliceDT{\multirput[bl]{0}(0,0)(10,10){5}}%
\def\aliceDX{\multirput[ur]{-90}(50,50)(10,-10){3}}%
\def\aliceDTHATCH{\aliceDT{\aliceDTSCALE{\pstickHATCH{\AliceColor}}}}%
\def\aliceDTLINE{\aliceDT{\aliceDTSCALE{\pstickLINE{\AliceColor}}}}%
\def\aliceDTCONE{\aliceDT{\aliceDTSCALE{\pstickCONE{\AliceColor}}}}%
\def\aliceDXHATCH{\aliceDX{\aliceDXSCALE{\pstickHATCH{\AliceColor!75}}}}%
\def\aliceDXLINE{\aliceDX{\aliceDXSCALE{\pstickLINE{\AliceColor!75}}}}%
\def\aliceDXCONE{\aliceDX{\aliceDXSCALE{\pstickCONE{\AliceColor!75}}}}%

\def\aliceDT{\multirput[bl]{0}(0,0)(10,10){5}}%
\def\aliceDTBHATCH{\aliceDT{\aliceDTSCALE{\pstickHATCH{\AliceColor}}}}%
\def\aliceDTBLINE{\aliceDT{\aliceDTSCALE{\pstickLINE{\AliceColor}}}}%
\def\aliceDTBCONE{\aliceDT{\aliceDTSCALE{\pstickCONE{\AliceColor}}}}%

\def\bobDTSCALE{\psscalebox{2 .5}}
\def\bobDXSCALE{\psscalebox{.5 2}}
\def\bobDTBSCALE{\psscalebox{2 .5}}
\def\bobDT{\multirput[bl]{0}(0,0)(20,5){6}}%
\def\bobDTB{\multirput[bl]{0}(40,-40)(20,5){6}}%
\def\bobDX{\multirput[ur]{-90}(0,0)(20,-5){5}}%
\def\bobDTHATCH{\bobDT{\bobDTSCALE{\pstickHATCH{\BobColor}}}}%
\def\bobDTLINE{\bobDT{\bobDTSCALE{\pstickLINE{\BobColor}}}}%
\def\bobDTCONE{\bobDT{\bobDTSCALE{\pstickCONE{\BobColor}}}}%
\def\bobDXHATCH{\bobDX{\bobDXSCALE{\pstickHATCH{\BobColor!75}}}}%
\def\bobDXLINE{\bobDX{\bobDXSCALE{\pstickLINE{\BobColor!75}}}}%
\def\bobDXCONE{\bobDX{\bobDXSCALE{\pstickCONE{\BobColor!75}}}}%

\def\bobDTBHATCH{\bobDTB{\bobDTBSCALE{\pstickHATCH{\BobColor}}}}%
\def\bobDTBLINE{\bobDTB{\bobDTBSCALE{\pstickLINE{\BobColor}}}}%
\def\bobDTBCONE{\bobDTB{\bobDTBSCALE{\pstickCONE{\BobColor}}}}%

\def\aliceMNX{\multirput[ur]{-90}(0,0)(10,-10){4}}%
\def\aliceMNXHATCH{\aliceMNX{\aliceDXSCALE{\pstickHATCH{black!75}}}}%
\def\aliceMNXLINE{\aliceMNX{\aliceDXSCALE{\pstickLINE{black!75}}}}%
\def\aliceMNXCONE{\aliceMNX{\aliceDXSCALE{\pstickCONE{black!75}}}}%

\bigskip

\begin{figure}[!Hht]
{%
\psset{unit=.85}
\begin{pspicture}(-50,0)(90,142)%
\psclip{\psframe[linestyle=none](-50,-0.25)(90,142)}{%
\rput[bl]{45}(0,0){%
\multirput[bl]{0}(0,-80)(0,10){25}{%
\multirput[bl]{0}(-50,0)(10,0){25}{\pstickSQUARE{black!75}}}}}%
\endpsclip%
\rput[bl]{45}(-42.426,28.284)
{%
\rput[bl]{0}(10,0){%
\rput[bl]{0}(0,0){\pstickDIAMOND{red}{40,-40}}%
\aliceDTHATCH%
\aliceDXHATCH%
\bobDTHATCH%
\bobDTBHATCH%
\bobDXHATCH%
\aliceMNXHATCH%
\aliceDTLINE%
\aliceDXLINE%
\bobDTLINE%
\bobDTBLINE%
\bobDXLINE%
\aliceMNXLINE%
\aliceDTCONE%
\aliceDXCONE%
\bobDTCONE%
\bobDXCONE%
\bobDTBCONE%
\aliceMNXCONE%
\pscircle(0,0){1}\rput[r]{-45}(0,0){$O\ $}%
\pscircle(50,50){1}\rput[r]{-45}(50,50){$P\ $}%
\pscircle(80,20){1}\rput[l]{-45}(80,20){$\ Q$}%
\pscircle(40,-40){1}\rput[l]{-45}(30,-40){$\ X $}%
\pscircle(100,-25){1}\rput[l]{-45}(100,-25){$\:\ Y$}%
\rput[b]{-45}(50,-12.5){\rput{30.9}(0,13){\psframebox[linecolor=black!50,framearc=.3]{\psframebox*[framearc=.3,fillcolor=white,opacity=.75]{\begin{tabular}{c}$L\BY{\BOB}{ladder}=5$\end{tabular}}}}}%
\rput[b]{-45}(20,-20){\rput(0,-13){\psframebox[linecolor=black!50,framearc=.3]{\psframebox*[framearc=.3,fillcolor=white,opacity=.75]{\begin{tabular}{c}$L\BY{\ALICE}{ladder}=4$\end{tabular}}}}}%

\rput[b]{-45}(100,30){\rput[r](-2,2){\psframebox*[boxsep=true,framesep=1pt,framearc=.3,fillcolor=white,opacity=.75]{\begin{tabular}{c}$\VEL\BY{\ALICE}{\Bob}=\displaystyle\frac{3}{5}\cspeed$\end{tabular}}}}%
\rput[b]{-45}(50,50){\rput[r](4,10){\psframebox*[framearc=.3,fillcolor=white,opacity=.75,framesep=1]{\bf Alice}}}%
\rput[b]{-45}(80,20){\rput[l](14,10){\psframebox*[framearc=.3,framesep=2,fillcolor=white,opacity=.75]{\bf Bob}}}%
\pscircle(50,-50){1}\rput[l]{-45}(50,-50){$\ Y' $}%
\pscircle(80,-20){1}\rput[b]{-45}(80,-20){\rput(0,7){$\:\ X'$}}%
}%
}%
\end{pspicture}%
}
\caption{\label{fig:LengthContraction} Length Contraction: $L\BY\ALICE{ladder} < L\BY\BOB{ladder} $.
}
\end{figure}
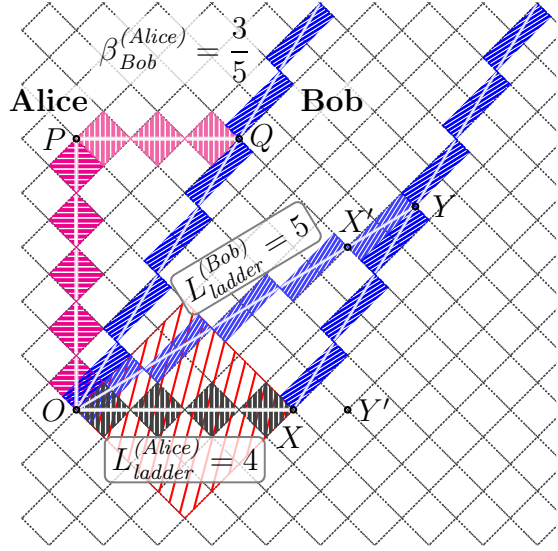 

\break

\subsection{Lorentz Boost coordinate transformation}
\label{sec:BoostTransformation}
{\it
Bob, traveling with velocity $(3/5)\cspeed$ according to Alice,
assigns event $E$ coordinates $(t\BY{\BOB}E,x\BY{\BOB}E)=(2,6)$.
What coordinates would Alice assign to $E$?
(Both Alice and Bob regard their meeting event $O$ as the origin of their coordinates.)
}

Refer to Fig.~\ref{fig:BoostTransformation}.
As before, we use Alice's diamonds 
to draw Bob's worldline and his diamonds. 
With Bob's diamonds, we locate the event $E$.
Now, using Alice's damonds, we find $(t\BY{\ALICE}E,x\BY{\ALICE}E)=(7,9)$.
This agrees with the result from the 
Lorentz Boost coordinate transformation formulae
(derived in Fig.~\ref{fig:BoostTransformationDerivation} and in the Appendix):
\begin{eqnarray}
t\BY{\ALICE}{E}
&=& 
\gamma\left( t\BY{\BOB}E + \overc{\VEL\BY{\ALICE}{\Bob}} x\BY{\BOB}E   \right) \label{eq:tBoost}
=\frac{5}{4} \left( 2 + \frac{3}{5} 6 \right)=7\\
x\BY{\ALICE}{E}
&=&\gamma \left( x\BY{\BOB}E + \overc{\VEL\BY{\ALICE}{\Bob}} t\BY{\BOB}E   \right) \label{eq:xBoost}
=\frac{5}{4} \left( 6 + \frac{3}{5} 2 \right)=9,
\end{eqnarray}
where $\gamma=5/4$ is the time-dilation factor (introduced in Sec.~\ref{sec:TimeDilation})
for velocity $\VEL\BY{\ALICE}{\Bob}=(3/5)\cspeed$.

In addition, note that the causal diamond with $OE$ as its spacelike diagonal
has area 32, which is equal to minus the square-interval of $OE$, 
$-((2)^2-(6)^2)= -( (7)^2-(9)^2)$.
We will elaborate on this point in Section~\ref{sec:Radar}.

\def\aliceDTSCALE{\psscalebox{1 1}}
\def\aliceDXSCALE{\psscalebox{1 1}}

\def\aliceDT{\multirput[bl]{0}(0,0)(10,10){7}}%
\def\aliceDX{\multirput[ur]{-90}(50,50)(10,-10){3}}%
\def\aliceDTHATCH{\aliceDT{\aliceDTSCALE{\pstickHATCH{\AliceColor}}}}%
\def\aliceDTLINE{\aliceDT{\aliceDTSCALE{\pstickLINE{\AliceColor}}}}%
\def\aliceDTCONE{\aliceDT{\aliceDTSCALE{\pstickCONE{\AliceColor}}}}%
\def\aliceDXHATCH{\aliceDX{\aliceDXSCALE{\pstickHATCH{\AliceColor!75}}}}%
\def\aliceDXLINE{\aliceDX{\aliceDXSCALE{\pstickLINE{\AliceColor!75}}}}%
\def\aliceDXCONE{\aliceDX{\aliceDXSCALE{\pstickCONE{\AliceColor!75}}}}%

\def\bobDTSCALE{\psscalebox{2 .5}}
\def\bobDXSCALE{\psscalebox{.5 2}}
\def\bobDTBSCALE{\psscalebox{2 .5}}
\def\bobDT{\multirput[bl]{0}(0,0)(20,5){4}}%
\def\bobDTB{\multirput[bl]{0}(40,-40)(20,5){4}}%
\def\bobDX{\multirput[ur]{-90}(40,10)(20,-5){6}}%
\def\bobDTHATCH{\bobDT{\bobDTSCALE{\pstickHATCH{\BobColor}}}}%
\def\bobDTLINE{\bobDT{\bobDTSCALE{\pstickLINE{\BobColor}}}}%
\def\bobDTCONE{\bobDT{\bobDTSCALE{\pstickCONE{\BobColor}}}}%
\def\bobDXHATCH{\bobDX{\bobDXSCALE{\pstickHATCH{\BobColor!75}}}}%
\def\bobDXLINE{\bobDX{\bobDXSCALE{\pstickLINE{\BobColor!75}}}}%
\def\bobDXCONE{\bobDX{\bobDXSCALE{\pstickCONE{\BobColor!75}}}}%

\def\bobDTBHATCH{\bobDTB{\bobDTBSCALE{\pstickHATCH{\BobColor}}}}%
\def\bobDTBLINE{\bobDTB{\bobDTBSCALE{\pstickLINE{\BobColor}}}}%
\def\bobDTBCONE{\bobDTB{\bobDTBSCALE{\pstickCONE{\BobColor}}}}%

\def\aliceMNX{\multirput[ur]{-90}(70,70)(10,-10){9}}%
\def\aliceMNXHATCH{\aliceMNX{\aliceDXSCALE{\pstickHATCH{black!75}}}}%
\def\aliceMNXLINE{\aliceMNX{\aliceDXSCALE{\pstickLINE{black!75}}}}%
\def\aliceMNXCONE{\aliceMNX{\aliceDXSCALE{\pstickCONE{black!75}}}}%

\begin{figure}[!Hht]
{%
\psset{unit=.85}
\begin{pspicture}(-50,0)(100,142)%
\psclip{\psframe[linestyle=none](-50,-0.25)(100,142)}{%
\rput[bl]{45}(0,0){%
\multirput[bl]{0}(0,-70)(0,10){25}{%
\multirput[bl]{0}(-50,0)(10,0){25}{\pstickSQUARE{black!75}}}}}%
\endpsclip%
\rput[bl]{45}(-42.426,14.142)
{%
\rput[bl]{0}(10,0){%
\rput[bl]{0}(0,0){\pstickDIAMOND{red}{160,-20}}%
\aliceMNXHATCH%
\aliceDTHATCH%
\aliceDXHATCH%
\bobDTHATCH%
\bobDXHATCH%
\aliceDTLINE%
\aliceDXLINE%
\bobDTLINE%
\bobDXLINE%
\aliceMNXLINE%
\aliceDTCONE%
\aliceDXCONE%
\bobDTCONE%
\bobDXCONE%
\aliceMNXCONE%
\pscircle(0,0){1}\rput[r]{-45}(0,0){$O\ $}%
\pscircle(70,70){1}\rput[r]{-45}(70,70){$T\ $}%
\pscircle(160,-20){1}\rput[l]{-45}(160,-20){$\ E $}%
\pscircle(40,10){1}\rput[r]{-45}(40,10){$T'\:\ $}%
\rput[b]{-45}(50,50){\rput[r](-5,0){\psframebox*[framearc=.3,fillcolor=white,opacity=.75,framesep=0.5]{\bf Alice}}}%
\rput[b]{-45}(80,20){\rput[l](6,0){\psframebox*[framearc=.3,fillcolor=white,opacity=.75,framesep=0.5]{\bf Bob}}}%
\rput[b]{-45}(160,-20){\rput[br](-2,10){\psframebox[linecolor=black!50,framearc=.3]{\psframebox*[framearc=.3,fillcolor=white,opacity=.75]{\begin{tabular}{c}$(t_E^{(A)},x_E^{(A)})=(7,9)$\end{tabular}}}}}%
\rput[b]{-45}(160,-20){\rput[tr]{30.9}(-2,-10){\psframebox[linecolor=black!50,framearc=.3]{\psframebox*[framearc=.3,fillcolor=white,opacity=.75]{\begin{tabular}{c}$(t_E^{(B)},x_E^{(B)})=(2,6)$\end{tabular}}}}}%
}%
}%
\end{pspicture}%
}
\caption{Comparison of coordinates without the Lorentz Boost coordinate transformation.
}
\label{fig:BoostTransformation} 
\end{figure}
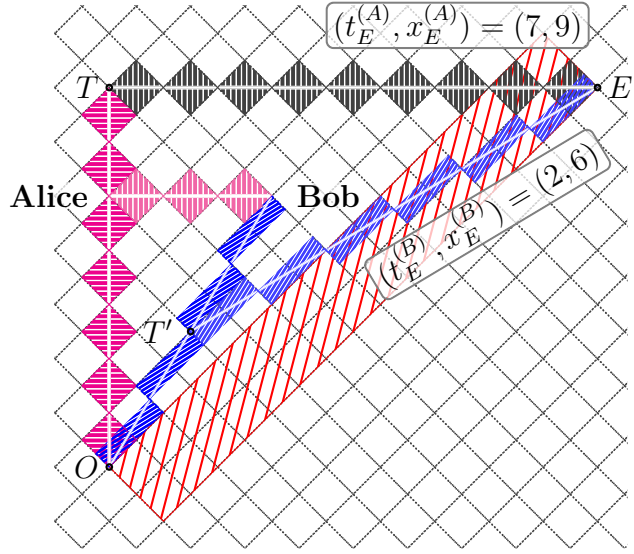

We can use the diagram to derive the inverse Lorentz Boost coordinate transformation.
Refer to Fig.~\ref{fig:BoostTransformationDerivation}. 
(A similar derivation  appears in Ellis.\cite{EllisWilliams})

\begin{figure}[!Hh]
{%
\psset{unit=.85}
\begin{pspicture}(-50,0)(100,142)%
\psclip{\psframe[linestyle=none](-50,-0.25)(100,142)}{%
\rput[bl]{45}(0,0){%
\multirput[bl]{0}(0,-70)(0,10){25}{%
\multirput[bl]{0}(-50,0)(10,0){25}{\pstickSQUARE{black!75}}}}}%
\endpsclip%
\rput[bl]{45}(-42.426,14.142)
{%
\rput[bl]{0}(10,0){%
\psline[linecolor=\AliceColor,linewidth=4]{c-c}(000,000)(070,070)
\psline[linecolor=\AliceColor,linewidth=4]{c-c}(050,050)(080,020)
\psline[linecolor=\BobColor,linewidth=4]{c-c}(000,000)(080,020)
\psline[linecolor=\BobColor,linewidth=4]{c-c}(040,010)(160,-020)
\psline[linecolor=black!75,linewidth=4]{c-c}(070,070)(160,-020)
\psline[linecolor=white,linewidth=2]{c-c}(040,010)(085,055)
\psline[linecolor=red,linewidth=.8]{c-c}(040,010)(085,055)
\psline[linecolor=white,linewidth=2]{c-c}(025,025)(070,070)
\psline[linecolor=red,linewidth=.8]{c-c}(025,025)(070,070)
\psline[linecolor=white,linewidth=2]{c-c}(160,-20)(085,055)
\psline[linecolor=red,linewidth=.8]{c-c}(160,-20)(085,055)
\psline[linecolor=white,linewidth=2]{c-c}(160,-20)(040,010)
\psline[linecolor=red,linewidth=.8]{c-c}(160,-20)(040,010)
\psline[linecolor=white,linewidth=2]{c-c}(040,010)(00,00)
\psline[linecolor=darkgreen,linewidth=.8]{c-c}(040,010)(00,00)
\psline[linecolor=white,linewidth=2]{c-c}(025,025)(00,00)
\psline[linecolor=darkgreen,linewidth=.8]{c-c}(025,025)(00,00)
\psline[linecolor=white,linewidth=2]{c-c}(040,010)(025,025)
\psline[linecolor=darkgreen,linewidth=.8]{c-c}(040,010)(025,025)
\psline[linecolor=white,linewidth=2]{c-c}(085,055)(070,070)
\psline[linecolor=darkgreen,linewidth=.8]{c-c}(085,055)(070,070)
\pscircle(50,50){1}\rput[r]{-45}(50,50){$P\ $}%
\pscircle(80,20){1}\rput[l]{-45}(80,20){$\:Q $}%
\pscircle(0,0){2}\pscircle[linecolor=white](0,0){1}\rput[r]{-45}(0,0){$O\ $}%
\pscircle(70,70){2}\pscircle[linecolor=white](70,70){1}\rput[r]{-45}(70,70){$T\ $}%
\pscircle(160,-20){2}\pscircle[linecolor=white](160,-20){1}\rput[l]{-45}(160,-20){$\ E $}%
\rput[b]{-45}(40,10){\rput[t](8,2){\psframebox*[framearc=.3,fillcolor=white,opacity=.75,framesep=0.5]{$T'$}}}%
\pscircle(40,10){2}\pscircle[linecolor=white](40,10){1}%
\pscircle[linecolor=black](25,25){1}\
\rput[r]{-45}(25,25){\rput[r](10,-4){\psframebox*[framearc=.3,fillcolor=white,opacity=.75,framesep=0.5]{$M$}}}%
\pscircle[linecolor=black](85,55){1}
\rput[b]{-45}(85,55){\rput[b](5,-10){\psframebox*[framearc=.3,fillcolor=white,opacity=.75,framesep=0.5]{$N$}}}%
\rput[b]{-45}(50,50){\rput[r](-5,12){\psframebox*[framearc=.3,fillcolor=white,opacity=.75,framesep=0.5]{\bf Alice}}}%
\rput[b]{-45}(80,20){\rput[l](6,12){\psframebox*[framearc=.3,fillcolor=white,opacity=.75,framesep=0.5]{\bf Bob}}}%
}%
}%
\rput[b](-25,25){
 \pscurve[linecolor=white,linewidth=4pt,arrowsize=2pt 2,arrowlength= 1.3,arrowinset=0,%
     showpoints=false]{->}(8,108)(3.3,107)(2.7,102)(2.7,98)(2.7,98)(2.7,32)
 \pscurve[linecolor=darkgreen,linewidth=2pt,%
     showpoints=false](8,108)(3.5,107)(2.7,102)(2.7,98)(2.7,98)(2.7,98)
 \pscurve[linecolor=darkgreen,linewidth=2pt,%
     showpoints=false]{->}(2.7,98)(2.7,98)(2.7,32)
 \pscurve[linecolor=white,linewidth=4pt,arrowsize=2pt 2,arrowlength= 1.3,arrowinset=0,%
     showpoints=false]{}(-30,51.3)(-29,55.6)(-20,57)(-20,57)
 \pscurve[linecolor=white,linewidth=4pt,arrowsize=2pt 2,arrowlength= 1.3,arrowinset=0,%
     showpoints=false]{}(-20,57)(-12,57)(10,57)
 \pscurve[linecolor=red,linewidth=2pt,arrowinset=0,%
     showpoints=false]{}(-30,51.3)(-29,55.6)(-20,57)(-20,57)
 \pscurve[linecolor=red,linewidth=2pt,arrowinset=0,%
     showpoints=false]{->}(-20,57)(-12,57)(10,57)
\rput[tc]{30.9}(14,014){\psframebox[linecolor=darkgreen!50,framearc=.3]{\psframebox*[framesep=1pt,framearc=.3,fillcolor=white,opacity=.75]{\boldmath $t_E^{(B)}$}}}%
\rput[tc]{0   }(-23,014){\psframebox[linecolor=darkgreen!50,framearc=.3]{\psframebox*[framesep=1pt,framearc=.3,fillcolor=white,opacity=.75]{\boldmath $\gamma t_E^{(B)}$}}}%
\rput[tc]{0   }(25, 108){\psframebox[linecolor=darkgreen!50,framearc=.3]{\psframebox*[framesep=1pt,framearc=.3,fillcolor=white,opacity=.75]{\boldmath $\VEL(\gamma t_E^{(B)})$}}}%
\rput[tc]{30.9}(62,052){\psframebox[linecolor=red!50,framearc=.3]{\psframebox*[framesep=1pt,framearc=.3,fillcolor=white,opacity=.75]{\boldmath $x_E^{(B)}$}}}%
\rput[tc]{0   }(62,108){\psframebox[linecolor=red!50,framearc=.3]{\psframebox*[framesep=1pt,framearc=.3,fillcolor=white,opacity=.75]{\boldmath $\gamma x_E^{(B)}$}}}%
\rput[tc]{0   }(-31, 45){\psframebox[linecolor=red!50,framearc=.3]{\psframebox*[framesep=1pt,framearc=.3,fillcolor=white,opacity=.75]{\boldmath $\VEL(\gamma x_E^{(B)})$}}}%
}%
\end{pspicture}%
}
\caption{A geometrical derivation of the Lorentz Boost coordinate transformation.
Since $\gamma=(OP)/(OQ)=(OM)/(OT')$ and $\gamma=(EN)/(ET')$ and
$\VEL=(PQ)/(OP)=(MT')/(OM)$ and $\VEL=(EN)/(NT')$, 
we have 
$t\BY{\ALICE}{E}=(OM)+(MT)=(OM)+(T'N)=(\gamma t\BY{\BOB}{E})+\VEL(\gamma x\BY{\BOB}{E})$
and
$x\BY{\ALICE}{E}=(TN)+(NE)=(MT')+(NE)=\VEL(\gamma t\BY{\BOB}{E})+\gamma x\BY{\BOB}{E}$.
A similar derivation appears in Ellis.\cite{EllisWilliams}
}
\label{fig:BoostTransformationDerivation} 
\end{figure}
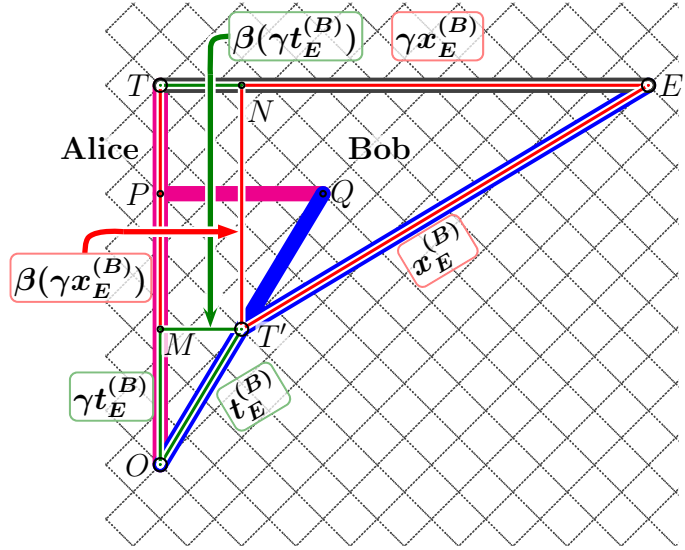

\subsection{Velocity Composition}
\label{sec:VelocityComposition}
{\it
Alice, Bob, and Carol met briefly at event $O$.
Bob, traveling with velocity $(3/5)\cspeed$ according to Alice,
observes Carol moving forward with velocity $(5/13)\cspeed$. 
What is the velocity of Carol according to Alice?}

Refer to Fig.~\ref{fig:VelocityComposition}.
Alice uses her diamonds to draw Bob's worldline and his diamonds,
with $OP$ and $PQ$ chosen so that $(PQ)/(OP)=3/5$.
Next, using Bob's diamonds, 
construct Carol's worldline along $OX$,
with $OX$ and $XY$ chosen so that $(XY)/(OX)=5/13$.
Alice, using her diamonds again,
determines the slope of Carol's worldline $OY$ 
by choosing an event $T$ on her worldline that she regards as simultaneous with $Y$.
Then, the velocity of Carol according to Alice is determined by $(TY)/(OT)$.
From the graph paper, we determine $\VEL\BYO{\ALICE}{\Carol}=(16/20)\cspeed=(4/5)\cspeed$.
This agrees with the result obtained by the velocity composition formula: 
\begin{eqnarray}
\VEL\BY{\ALICE}{\Carol}
&=&\frac{\VEL\BY{\ALICE}{\Bob}+\VEL\BY{\BOB}{\Carol}}{1+\overcc{\VEL\BY{\ALICE}{\Bob} \VEL\BY{\BOB}{\Carol}} } \label{eq:velcomp}\\
&=&\frac{\frac{3}{5} + \frac{5}{13}}{1+\left(\frac{3}{5}\right)\left(\frac{5}{13}\right)}\cspeed=\frac{4}{5}\cspeed \nonumber
\end{eqnarray}
(Proof: Write $\VEL\BY{\ALICE}{\Carol}=x\BY{\ALICE}{Y}/t\BY{\ALICE}{Y}$, 
then substitute the analogues of Eq.~(\ref{eq:xBoost}) and Eq.~(\ref{eq:tBoost}). 
Since $x\BY{\BOB}{Y}=\VEL\BY{\BOB}{\Carol}t\BY{\BOB}{Y}$, the velocity composition formula follows.
A similar derivation appears in Taylor.\cite{TaylorWheeler} %
An alternate proof based on radar-methods is given in the Appendix.)

\def\aliceDTSCALE{\psscalebox{1 1}}
\def\aliceDXSCALE{\psscalebox{1 1}}

\def\aliceDT{\multirput[bl]{0}(0,0)(10,10){20}}%
\def\aliceDX{\multirput[ur]{-90}(50,50)(10,-10){3}}%
\def\aliceDTHATCH{\aliceDT{\aliceDTSCALE{\pstickHATCH{\AliceColor}}}}%
\def\aliceDTLINE{\aliceDT{\aliceDTSCALE{\pstickLINE{\AliceColor}}}}%
\def\aliceDTCONE{\aliceDT{\aliceDTSCALE{\pstickCONE{\AliceColor}}}}%
\def\aliceDXHATCH{\aliceDX{\aliceDXSCALE{\pstickHATCH{\AliceColor!75}}}}%
\def\aliceDXLINE{\aliceDX{\aliceDXSCALE{\pstickLINE{\AliceColor!75}}}}%
\def\aliceDXCONE{\aliceDX{\aliceDXSCALE{\pstickCONE{\AliceColor!75}}}}%

\def\bobDTSCALE{\psscalebox{2 .5}}
\def\bobDXSCALE{\psscalebox{.5 2}}
\def\bobDTBSCALE{\psscalebox{2 .5}}
\def\bobDT{\multirput[bl]{0}(0,0)(20,5){13}}%
\def\bobDTB{\multirput[bl]{0}(40,-40)(20,5){6}}%
\def\bobDX{\multirput[ur]{-90}(260,65)(20,-5){5}}%
\def\bobDTHATCH{\bobDT{\bobDTSCALE{\pstickHATCH{\BobColor}}}}%
\def\bobDTLINE{\bobDT{\bobDTSCALE{\pstickLINE{\BobColor}}}}%
\def\bobDTCONE{\bobDT{\bobDTSCALE{\pstickCONE{\BobColor}}}}%
\def\bobDXHATCH{\bobDX{\bobDXSCALE{\pstickHATCH{\BobColor!75}}}}%
\def\bobDXLINE{\bobDX{\bobDXSCALE{\pstickLINE{\BobColor!75}}}}%
\def\bobDXCONE{\bobDX{\bobDXSCALE{\pstickCONE{\BobColor!75}}}}%

\def\bobDTBHATCH{\bobDTB{\bobDTBSCALE{\pstickHATCH{\BobColor}}}}%
\def\bobDTBLINE{\bobDTB{\bobDTBSCALE{\pstickLINE{\BobColor}}}}%
\def\bobDTBCONE{\bobDTB{\bobDTBSCALE{\pstickCONE{\BobColor}}}}%

\def\aliceMNX{\multirput[ur]{-90}(200,200)(10,-10){16}}%
\def\aliceMNXHATCH{\aliceMNX{\aliceDXSCALE{\pstickHATCH{black!75}}}}%
\def\aliceMNXLINE{\aliceMNX{\aliceDXSCALE{\pstickLINE{black!75}}}}%
\def\aliceMNXCONE{\aliceMNX{\aliceDXSCALE{\pstickCONE{black!75}}}}%

\def\carolDTSCALE{\psscalebox{3 .333}}
\def\carolDXSCALE{\psscalebox{.333 3}}
\def\carolDTBSCALE{\psscalebox{3 .333}}
\def\carolDT{\multirput[bl]{0}(0,0)(30,3.333){12}}%
\def\carolDTB{\multirput[bl]{0}(40,-40)(20,5){6}}%
\def\carolDX{\multirput[ur]{-90}(260,65)(20,-5){5}}%
\def\carolDTHATCH{\carolDT{\carolDTSCALE{\pstickHATCH{\CarolColor}}}}%
\def\carolDTLINE{\carolDT{\carolDTSCALE{\pstickLINE{\CarolColor}}}}%
\def\carolDTCONE{\carolDT{\carolDTSCALE{\pstickCONE{\CarolColor}}}}%
\def\carolDTLINEBOLD{\carolDT{\carolDTSCALE{\pstickLINEBOLD{\CarolColor}}}}%

\begin{figure}[!Hht]
{%
\psset{unit=.5}
\begin{pspicture}(-70,0)(210,302)%
\psclip{\psframe[linestyle=none](-70,-0.25)(210,302)}{%
\rput[bl]{45}(0,0){%
\multirput[bl]{0}(0,-180)(0,10){50}{%
\multirput[bl]{0}(-50,0)(10,0){50}{\pstickSQUARE{black!75}}}}}%
\endpsclip%
\rput[bl]{45}(-42.426,0)
{%
\rput[bl]{0}(10,0){%
\aliceMNXHATCH%
\aliceDTHATCH%
\aliceDXHATCH%
\bobDTHATCH%
\bobDXHATCH%
\aliceDTLINE%
\aliceDXLINE%
\bobDTLINE%
\bobDXLINE%
\aliceMNXLINE%
\aliceDTCONE%
\aliceDXCONE%
\bobDTCONE%
\bobDXCONE%
\aliceMNXCONE%
\pscircle(0,0){1}\rput[r]{-45}(0,0){$O\ $}%
\pscircle(50,50){1}\rput[r]{-45}(50,50){$P\ $}%
\pscircle(80,20){1}\rput[l]{-45}(80,20){$\:Q $}%
\pscircle(260,65){1}\rput[r]{-45}(260,65){$X\:\ $}%
\pscircle(360,40){1}\rput[l]{-45}(360,40){$\:Y$}%
\pscircle(200,200){1}\rput[r]{-45}(200,200){$T\ $}%
\psline[linecolor=\CarolColor,linewidth=.6]{c-c}(360,40)(0,0)
}%
}%
\rput[l](-28,200){\rput[l]{0}(0,0){\psframebox*[framearc=.3,fillcolor=white,opacity=.75,framesep=1]{\bf Alice}}}
\rput[l](44,200){\rput[l]{0}(0,0){\psframebox*[framearc=.3,fillcolor=white,opacity=.75,framesep=1]{\bf Bob}}}
\rput[l](125,200){\rput[l]{0}(0,0){\psframebox*[framearc=.3,fillcolor=white,opacity=.75,framesep=1]{\bf Carol}}}
\end{pspicture}%
}
\caption{\label{fig:VelocityComposition} Velocity Composition
}
\end{figure}
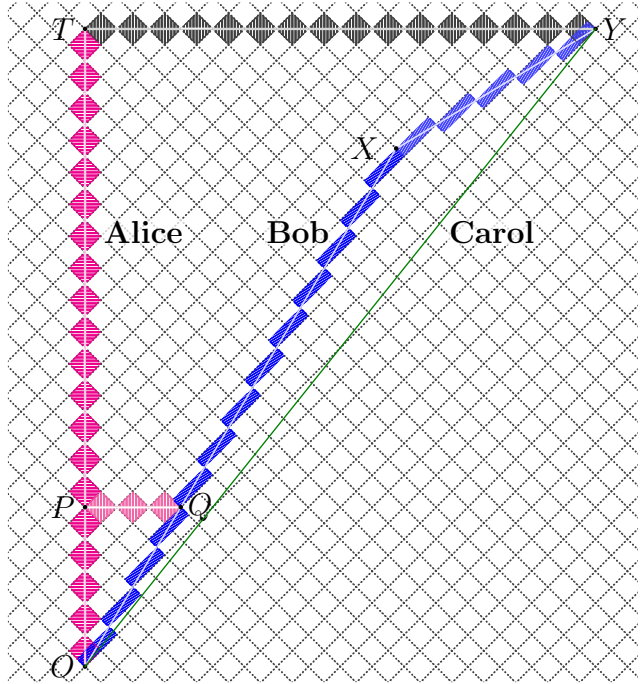

\subsection{Radar Measurements and the Invariant Square-Interval}
\label{sec:Radar}

\def\aliceDTSCALE{\psscalebox{1 1}}
\def\aliceDXSCALE{\psscalebox{1 1}}

\def\aliceDT{\multirput[bl]{0}(0,0)(10,10){20}}%
\def\aliceDX{\multirput[ur]{-90}(50,50)(10,-10){3}}%
\def\aliceDTHATCH{\aliceDT{\aliceDTSCALE{\pstickHATCH{\AliceColor}}}}%
\def\aliceDTLINE{\aliceDT{\aliceDTSCALE{\pstickLINE{\AliceColor}}}}%
\def\aliceDTCONE{\aliceDT{\aliceDTSCALE{\pstickCONE{\AliceColor}}}}%
\def\aliceDXHATCH{\aliceDX{\aliceDXSCALE{\pstickHATCH{\AliceColor!75}}}}%
\def\aliceDXLINE{\aliceDX{\aliceDXSCALE{\pstickLINE{\AliceColor!75}}}}%
\def\aliceDXCONE{\aliceDX{\aliceDXSCALE{\pstickCONE{\AliceColor!75}}}}%

\def\bobDTSCALE{\psscalebox{2 .5}}
\def\bobDXSCALE{\psscalebox{.5 2}}
\def\bobDTBSCALE{\psscalebox{2 .5}}
\def\bobDT{\multirput[bl]{0}(0,0)(20,5){16}}%
\def\bobDTB{\multirput[bl]{0}(40,-40)(20,5){6}}%
\def\bobDX{\multirput[ur]{-90}(260,65)(20,-5){5}}%
\def\bobDTHATCH{\bobDT{\bobDTSCALE{\pstickHATCH{\BobColor}}}}%
\def\bobDTLINE{\bobDT{\bobDTSCALE{\pstickLINE{\BobColor}}}}%
\def\bobDTCONE{\bobDT{\bobDTSCALE{\pstickCONE{\BobColor}}}}%
\def\bobDXHATCH{\bobDX{\bobDXSCALE{\pstickHATCH{\BobColor!75}}}}%
\def\bobDXLINE{\bobDX{\bobDXSCALE{\pstickLINE{\BobColor!75}}}}%
\def\bobDXCONE{\bobDX{\bobDXSCALE{\pstickCONE{\BobColor!75}}}}%

\def\bobDTBHATCH{\bobDTB{\bobDTBSCALE{\pstickHATCH{\BobColor}}}}%
\def\bobDTBLINE{\bobDTB{\bobDTBSCALE{\pstickLINE{\BobColor}}}}%
\def\bobDTBCONE{\bobDTB{\bobDTBSCALE{\pstickCONE{\BobColor}}}}%

\def\aliceMNX{\multirput[ur]{-90}(200,200)(10,-10){16}}%
\def\aliceMNXHATCH{\aliceMNX{\aliceDXSCALE{\pstickHATCH{black!75}}}}%
\def\aliceMNXLINE{\aliceMNX{\aliceDXSCALE{\pstickLINE{black!75}}}}%
\def\aliceMNXCONE{\aliceMNX{\aliceDXSCALE{\pstickCONE{black!75}}}}%

\def\aliceDAVIDX{\multirput[ur]{-90}(130,130)(10,-10){5}}%
\def\aliceDAVIDXHATCH{\aliceDAVIDX{\aliceDXSCALE{\pstickHATCH{\AliceColor!75}}}}%
\def\aliceDAVIDXLINE{\aliceDAVIDX{\aliceDXSCALE{\pstickLINE{\AliceColor!75}}}}%
\def\aliceDAVIDXCONE{\aliceDAVIDX{\aliceDXSCALE{\pstickCONE{\AliceColor!75}}}}%

\def\carolDTSCALE{\psscalebox{3 .333}}
\def\carolDXSCALE{\psscalebox{.333 3}}
\def\carolDTBSCALE{\psscalebox{3 .333}}
\def\carolDT{\multirput[bl]{0}(0,0)(30,3.333){12}}%
\def\carolDTB{\multirput[bl]{0}(40,-40)(30,3.333){6}}%
\def\carolDX{\multirput[ur]{-90}(260,65)(30,-3.333){5}}%
\def\carolDTHATCH{\carolDT{\carolDTSCALE{\pstickHATCH{\CarolColor}}}}%
\def\carolDTLINE{\carolDT{\carolDTSCALE{\pstickLINE{\CarolColor}}}}%
\def\carolDTCONE{\carolDT{\carolDTSCALE{\pstickCONE{\CarolColor}}}}%
\def\carolDTLINEBOLD{\carolDT{\carolDTSCALE{\pstickLINEBOLD{\CarolColor}}}}%

\def\davidDTSCALE{\psscalebox{1.5 .666}}
\def\davidDXSCALE{\psscalebox{.666 1.5}}
\def\davidDTBSCALE{\psscalebox{1.5 .666}}
\def\davidDT{\multirput[bl]{0}(0,0)(15,6.666){15}}%
\def\davidDTB{\multirput[bl]{0}(40,-40)(15,6.666){6}}%
\def\davidDX{\multirput[ur]{-90}(225,100)(15,-6.666){9}}%
\def\davidDTHATCH{\davidDT{\davidDTSCALE{\pstickHATCH{\DavidColor}}}}%
\def\davidDTLINE{\davidDT{\davidDTSCALE{\pstickLINE{\DavidColor}}}}%
\def\davidDTCONE{\davidDT{\davidDTSCALE{\pstickCONE{\DavidColor}}}}%
\def\davidDXHATCH{\davidDX{\davidDXSCALE{\pstickHATCH{\DavidColor!75}}}}%
\def\davidDXLINE{\davidDX{\davidDXSCALE{\pstickLINE{\DavidColor!75}}}}%
\def\davidDXCONE{\davidDX{\davidDXSCALE{\pstickCONE{\DavidColor!75}}}}%
\def\davidDTLINEBOLD{\davidDT{\davidDTSCALE{\pstickLINEBOLD{\DavidColor}}}}%

\begin{figure}[!Hht]
{%
\psset{unit=.5}
\begin{pspicture}(-70,0)(210,302)%
\psclip{\psframe[linestyle=none](-70,-0.25)(210,302)}{%
\rput[bl]{45}(0,0){%
\multirput[bl]{0}(0,-180)(0,10){50}{%
\multirput[bl]{0}(-50,0)(10,0){50}{\pstickSQUARE{black!75}}}}}%
\endpsclip%
\rput[bl]{45}(-42.426,0)
{%
\rput[bl]{0}(10,0){%
\rput[bl]{0}(0,0){\pstickDIAMOND{red}{180,20}}%
\aliceMNXHATCH%

\aliceDTHATCH%
\aliceDXHATCH%
\bobDTHATCH%
\carolDTHATCH%
\aliceDTLINE%
\aliceDXLINE%
\bobDTLINE%
\aliceMNXLINE%
\aliceDTCONE%
\aliceDXCONE%
\bobDTCONE%
\aliceMNXCONE%
\carolDTLINE%
\rput[bl]{0}(20,20){\pstickLIGHTLINE{black}{160,00}}%
\rput[bl]{0}(180,20){\pstickLIGHTLINE{black}{0,160}}%
\pscircle(0,0){2}\pscircle[linecolor=white](0,0){1}\rput[r]{-45}(0,0){$O\ $}%
\pscircle(180,20){2}\pscircle[linecolor=white](180,20){1}\rput[l]{-45}(180,20){$\ W$}%
\pscircle(20,20){2}\pscircle[linecolor=white](20,20){1}\rput[r]{-45}(20,20){$r\ $}%
\pscircle(180,180){2}\pscircle[linecolor=white](180,180){1}\rput[r]{-45}(180,180){$R\ $}%
\pscircle(80,20){2}\pscircle[linecolor=white](80,20){1}\rput[l]{-45}(80,20){\rput[l](-4,12){\psframebox*[framearc=.3,fillcolor=white,opacity=0,framesep=0.5]{$s $}}}%
\pscircle(180,45){2}\pscircle[linecolor=white](180,45){1}\rput[c]{-45}(180,45){\rput[l](-15,-2){$S\ $}}%
}%
}%
\rput[l](-28,200){\rput[l]{0}(0,0){\psframebox*[framearc=.3,fillcolor=white,opacity=.75,framesep=1]{\bf Alice}}}
\rput[l](44,200){\rput[l]{0}(0,0){\psframebox*[framearc=.3,fillcolor=white,opacity=.75,framesep=1]{\bf Bob}}}
\rput[l](125,200){\rput[l]{0}(0,0){\psframebox*[framearc=.3,fillcolor=white,opacity=.75,framesep=1]{\bf Carol}}}
\end{pspicture}%
}
\caption{\label{fig:RadarMeasurements} Radar Measurements of event~$W$.
}
\end{figure}
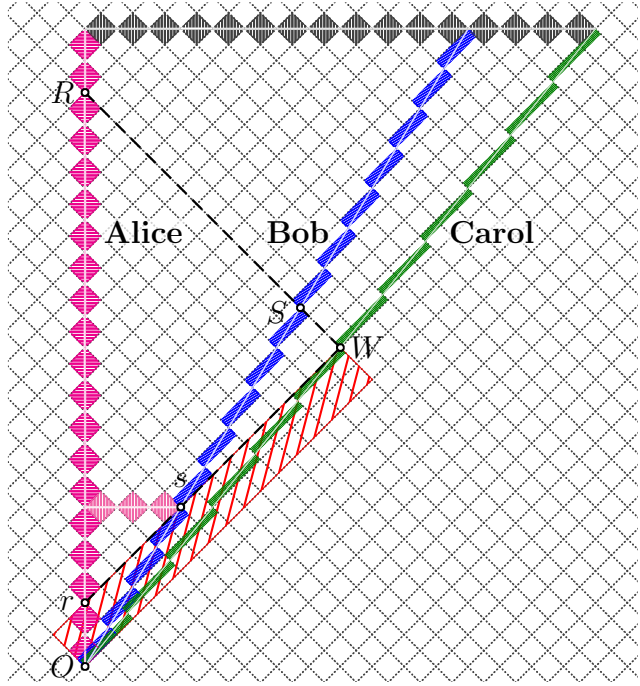

We reuse parts of the previous situation 
in Sec.~\ref{sec:VelocityComposition}
to describe radar measurements.
(Refer to Fig.~\ref{fig:RadarMeasurements}.)
In this section, note the many interesting kinematical quantities
that can be
expressed in terms of the counting of clock ticks 
on an observer's worldline.%
\cite{Robb, Marzke,Operational, Bondi, SyngeGR, Geroch, EllisWilliams, Woodhouse}
 We feel that this is
a more natural starting point for the development 
of the subject.

Alice operationally assigns coordinates to event $W$ using a radar-measurement,
emitting a light-signal at event $r$ to the target event $W$, then receiving its echo at event $R$.
From the two elapsed times (since the origin event $O$) on her worldline, $t\BY{\ALICE}{\vphantom{R}r}$ and $t\BY{\ALICE}R$. 
Alice's rectangular coordinates for $W$ are obtained by
\begin{equation}
\label{eq:rectCoordsA}
\mbox{
$t\BY{\ALICE}W=\frac{1}{2}(t\BY{\ALICE}R + t\BY{\ALICE}{\vphantom{R}r})$
\qquad
and
\qquad 
$x\BY{\ALICE}W=\frac{1}{2}(t\BY{\ALICE}R - t\BY{\ALICE}{\vphantom{R}r})$,
}
\end{equation}
which can be interpreted as the ``midway time between emission and reception''
and ``half of the round-trip distance,'' respectively.
From the diagram, Alice measures $t\BY{\ALICE}{\vphantom{R}r}=2$ and $t\BY{\ALICE}{R}=18$, so that
$t\BY{\ALICE}W=10$ and $x\BY{\ALICE}W=8$.
Similarly, for Bob, using events $s$ and $S$,
\begin{equation}
\label{eq:rectCoordsB}
\mbox{
$t\BY{\BOB}W=\frac{1}{2}(t\BY{\BOB}S + t\BY{\BOB}{\vphantom{S}s})$
\qquad
and
\qquad 
$x\BY{\BOB}W=\frac{1}{2}(t\BY{\BOB}S - t\BY{\BOB}{\vphantom{S}s})$,
}
\end{equation}
with
$t\BY{\BOB}{\vphantom{S}s}=4$ and $t\BY{\BOB}S=9$, so that
$t\BY{\BOB}W=6.5$ and $x\BY{\BOB}W=2.5$.

In general, Alice and Bob disagree on their $t$- and $x$-coordinates for $W$.
However, they agree on the invariant square-interval
$(t_W)^2 - (x_W)^2$, which can be expressed as the
product of their radar times:\cite{Robb,SyngeGR,Geroch}
\begin{equation}
(t_W)^2 - (x_W)^2=t\BY{\ALICE}R t\BY{\ALICE}{\vphantom{R}r}=t\BY{\BOB}S t\BY{\BOB}{\vphantom{S}s}=36.\label{eq:SquareInterval}
\end{equation}

Let us observe that, 
\textit{for each inertial observer, the pair of radar times correspond 
to that observer's light-cone coordinates\cite{DiracCoords} of event $W$}.
From Eq.~(\ref{eq:LightConeCoords}), we have $u\equiv t+x$ and $v\equiv t-x$.
Since events $r$ and $R$ are on Alice's worldline (where $x\BY\ALICE{r}=0$), 
$u\BY{\ALICE}{\vphantom{R}r}=v\BY{\ALICE}{\vphantom{R}r}=t\BY{\ALICE}{\vphantom{R}r}$ and 
$u\BY{\ALICE}{R}=v\BY{\ALICE}{R}=t\BY{\ALICE}{R}$. 
Refer again to Fig.~\ref{fig:RadarMeasurements}.
Since $W$ is on the past-light-cone of event~$R$ and $x\BY{\ALICE}W>0$, 
we have $u\BY{\ALICE}{W}=u\BY{\ALICE}{R}$.
Since $W$ is on the future-light-cone of event~$r$ and $x\BY{\ALICE}W>0$, 
we have $v\BY{\ALICE}{W}=v\BY{\ALICE}{\vphantom{R}r}$.
Thus, 
\begin{equation}
(u\BY{\ALICE}{W},v\BY{\ALICE}{W})=(t\BY{\ALICE}{R},t\BY{\ALICE}{\vphantom{R}r}).\label{eq:LCAlice}
\end{equation}
Similarly, 
\begin{equation}
(u\BY{\BOB}{W},v\BY{\BOB}{W})=(t\BY{\BOB}{S},t\BY{\BOB}{\vphantom{S}s}).\label{eq:LCBob}
\end{equation}

Furthermore,
this means that the invariant square-interval 
in Eq.~(\ref{eq:SquareInterval}) can be written 
in light-cone coordinates as
\begin{equation}
u\BY{\ALICE}{W}v\BY{\ALICE}{W}=u\BY{\BOB}{W}v\BY{\BOB}{W}=36,
\label{eq:areaLCcoords}
\end{equation} 
which is equal to the \textit{area of the causal diamond of $OW$}.
For Carol (along $OW$), her radar experiment for $W$
has her emission and reception events coinciding with the target event $W$. 
Thus, $(t\BY{\CAROL}W)^2=36$, which means that there are $\sqrt{36}=6$ of Carol's 
light-clock diamonds
from event~$O$ to event~$W$.

From the rectangular coordinates, one can operationally 
determine the relative velocities of the inertial observers
by, for example,
\begin{equation}
\VEL\BY{\ALICE}{\Carol}
=\frac{x\BY{\ALICE}{W}-x\BY{\ALICE}{O}}{t\BY{\ALICE}{W}-t\BY{\ALICE}{O}}
=\frac{(t\BY{\ALICE}R-t\BY{\ALICE}{\vphantom{R}r})-0}{(t\BY{\ALICE}R+t\BY{\ALICE}{\vphantom{R}r})-0}
.
\end{equation}
This velocity can be related to the
$k$-factors first introduced in Sec.~\ref{sec:BobLightClock}.
Applying the idea behind Eq.~(\ref{eq:Ksq}),
we have
$
(k_{\Alice,\Carol})^2 = t\BY{\ALICE}R/t\BY{\ALICE}{\vphantom{R}r}
$. 
Thus,
\begin{equation}
\label{eq:vAC}
\VEL\BY{\ALICE}{\Carol}
=\frac{\left(\frac{(t\BY{\ALICE}R)}{t\BY{\ALICE}{\vphantom{R}r}}\right)-1}{\left(\frac{(t\BY{\ALICE}R)}{t\BY{\ALICE}{\vphantom{R}r}}\right)+1}
=\frac{(k_{\Alice,\Carol})^2-1}{(k_{\Alice,\Carol})^2+1}.
\end{equation}
Upon solving that equation for 
$k_{\Alice,\Carol}$, 
we can recognize it as the familiar Doppler factor 
\begin{equation}
\label{eq:kvel}
k_{\Alice,\Carol}
=\sqrt{\frac{ 1 + \VEL\BY{\ALICE}{\Carol} }{ 1 - \VEL\BY{\ALICE}{\Carol} }}
,
\end{equation}
which we had given earlier (without proof) as Eq.~(\ref{eq:BondiKfactor}).

With Alice's measurements of event~$W$ obtained earlier,
$t\BY{\ALICE}{\vphantom{R}r}=2$ and $t\BY{\ALICE}{R}=18$,
we have 
$k_{\Alice,\Carol}=\sqrt{ t\BY{\ALICE}{R} / t\BY{\ALICE}{\vphantom{R}r} }=\sqrt{18/2}=3$ 
and 
$\VEL\BY{\ALICE}{\Carol}=(18-2)/(18+2)=4/5$.
Similarly, 
with Bob's measurements of $t\BY{\BOB}{\vphantom{S}s}=4$ and $t\BY{\BOB}{S}=9$,
we have 
$k_{\Bob,\Carol}=\sqrt{t\BY{\BOB}{S} / t\BY{\BOB}{\vphantom{S}s} }=\sqrt{9/4}=3/2$ and 
$\VEL\BY{\BOB}{\Carol}=(9-4)/(9+4)=5/13$.
These are consistent with the velocity composition analysis
in Sec.~\ref{sec:VelocityComposition}.

At this point, one could go further in this viewpoint 
to derive the velocity composition formula and the 
Lorentz Boost coordinate transformations.
We do this in the Appendix.%

\subsection{Collisions in Energy-Momentum Space}
\label{sec:EnergyMomentum}

\def\EMpstickHATCH#1{%
\psframe[linecolor={#1},dimen=middle,linestyle=none,linewidth=.1pt,fillstyle=vlines,%
hatchcolor={#1},hatchsep=1.100,hatchwidth=.3142,addfillstyle=vlines,opacity=0.5](0,0)(10,10)%
}%

\def\allyDTSCALE{\psscalebox{1 1 }}
\def\allyDXSCALE{\psscalebox{1 1 }}
\def\allyDT{\multirput[bl]{0}(0,0)(10, 10){13}}%
\def\allyDX{\multirput[ur]{90}(130,130)(-10,10){5}}%
\def\allyDTHATCH{\allyDT{\allyDTSCALE{\EMpstickHATCH{\AliceColor}}}}%
\def\allyDTWLINE{\allyDT{\allyDTSCALE{\pstickLINE{\AliceColor}}}}%
\def\allyDTLCONE{\allyDT{\allyDTSCALE{\pstickCONE{\AliceColor}}}}%
\def\allyDXHATCH{\allyDX{\allyDXSCALE{\EMpstickHATCH{\AliceColor!75}}}}%
\def\allyDXLINE{\allyDX{\allyDXSCALE{\pstickLINE{\AliceColor!75}}}}%
\def\allyDXCONE{\allyDX{\allyDXSCALE{\pstickCONE{\AliceColor!75}}}}%
\def\massADTSCALE{\psscalebox{.666 1.5}}
\def\massADT{\multirput[bl]{0}(0,0)(6.66,15){12}}%
\def\massADTBSCALE{\psscalebox{3 .333}}
\def\massADTB{\multirput[bl]{0}(0,0)(30,3.33){12}}%
\def\massADTHATCH{\massADT{\massADTSCALE{\EMpstickHATCH{\BobColor}}}}%
\def\massADTWLINE{\massADT{\massADTSCALE{\pstickLINE{\BobColor}}}}%
\def\massADTLCONE{\massADT{\massADTSCALE{\pstickCONE{\BobColor}}}}%
\def\massADTBHATCH{\massADTB{\massADTBSCALE{\EMpstickHATCH{\BobColor}}}}%
\def\massADTBWLINE{\massADTB{\massADTBSCALE{\pstickLINE{\BobColor}}}}%
\def\massADTBLCONE{\massADTB{\massADTBSCALE{\pstickCONE{\BobColor}}}}%

\def\massBDTSCALE{\psscalebox{4 .25}}
\def\massBDTBSCALE{\psscalebox{.5 2}}
\def\massBDT{\multirput[bl]{0}(80,180)(40,2.5){8}}%
\def\massBDTB{\multirput[bl]{0}(360,40)(5,20){8}}%
\def\massBDX{\multirput[ur]{-90}(260,65)(30,-3.333){5}}%
\def\massBDTHATCH{\massBDT{\massBDTSCALE{\EMpstickHATCH{\CarolColor}}}}%
\def\massBDTWLINE{\massBDT{\massBDTSCALE{\pstickLINE{\CarolColor}}}}%
\def\massBDTLCONE{\massBDT{\massBDTSCALE{\pstickCONE{\CarolColor}}}}%
\def\massBDTWLINEBOLD{\massBDT{\massBDTSCALE{\pstickLINEBOLD{\CarolColor}}}}%
\def\massBDTBHATCH{\massBDTB{\massBDTBSCALE{\EMpstickHATCH{\CarolColor}}}}%
\def\massBDTBWLINE{\massBDTB{\massBDTBSCALE{\pstickLINE{\CarolColor}}}}%
\def\massBDTBLCONE{\massBDTB{\massBDTBSCALE{\pstickCONE{\CarolColor}}}}%
\def\massBDTBWLINEBOLD{\massBDTB{\massBDTBSCALE{\pstickLINEBOLD{\CarolColor}}}}%

\begin{figure}[!Hh]
{%
\psset{unit=.33}
\begin{pspicture}(-180,0)(250,460)%
\psclip{\psframe[linestyle=none](-180,-0.25)(250,455)}{%
\rput[bl]{45}(0,0){%
\multirput[bl]{0}(0,-180)(0,10){150}{%
\multirput[bl]{0}(-120,0)(10,0){70}{\pstickSQUARE{black!75}}}}}%
\endpsclip%
\rput[l](-125,120){\rput[l]{0}(0,0){\psframebox*[framearc=.3,fillcolor=white,opacity=.75,framesep=1]{ $\vec P_{1,i}$}}} %
\rput[l](45,170){\rput[l]{0}(0,0){\psframebox*[framearc=.3,fillcolor=white,opacity=.75,framesep=1]{   $\vec P_{1,f}$}}}
\rput[r](-10,290){\rput[l]{0}(0,0){\psframebox*[framearc=.3,fillcolor=white,opacity=.75,framesep=1]{  $\vec P_{2,i}$}}}
\rput[l](103,352){\rput[l]{0}(0,0){\psframebox*[framearc=.3,fillcolor=white,opacity=.75,framesep=1]{  $\vec P_{2,f}$}}}
\rput[b](100,424){\rput[b]{0}(20,6){\psframebox*[framearc=.3,fillcolor=white,opacity=.75,framesep=1]{ $\vec P_{COM}$}}}
\rput[bl]{45}(-42.426,0)
{%
\rput[bl]{0}(10,0){%
\psline[linecolor=\CarolColor,linewidth=.6]{c-c}(400,200)(0,0)
\psline[linecolor=red,linewidth=.8]{c-c}(000,000)(000,200)
\psline[linecolor=red,linewidth=.8]{c-c}(000,200)(400,200)
\psline[linecolor=red,linewidth=.8]{c-c}(400,200)(400,000)
\psline[linecolor=red,linewidth=.8]{c-c}(400,000)(0,0)
\allyDTHATCH%
\allyDXHATCH%
\massADTHATCH%
\massADTBHATCH%
\massBDTHATCH%
\massBDTBHATCH%
\allyDTWLINE%
\allyDXLINE%
\massADTWLINE%
\massADTBWLINE%

\allyDTLCONE%
\allyDXCONE%
\massADTLCONE%
\massADTBLCONE%
\massBDTLCONE%
\massBDTBLCONE%
\massBDTWLINE%
\massBDTBWLINE%

\pscircle(0,0){2}\pscircle[linecolor=white](0,0){1}\rput[r]{-45}(0,0){$O\ $}%
\pscircle(400,200){2}\pscircle[linecolor=white](400,200){1}%
}%
}%
\end{pspicture}%
}
\caption{\label{fig:EnergyMomentum} 
An elastic collision (with {%
$\vec P_{1,i}+\vec P_{2,i}=\vec P_{1,f}+\vec P_{2,f}$})
drawn on Alice's energy-momentum diagram.
The particles have rest-masses $m_1=12$ and $m_2=8$,
initial-velocities $\VEL_{1,i}=-5/13$ and $\VEL_{2,i}=15/17$,
and final-velocities $\VEL_{1,f}=4/5$ and $\VEL_{2,f}=-3/5$.
We have drawn a ``mass diamond'' for the total momentum vector, {%
$\vec P_{COM}$}.
}
\end{figure}
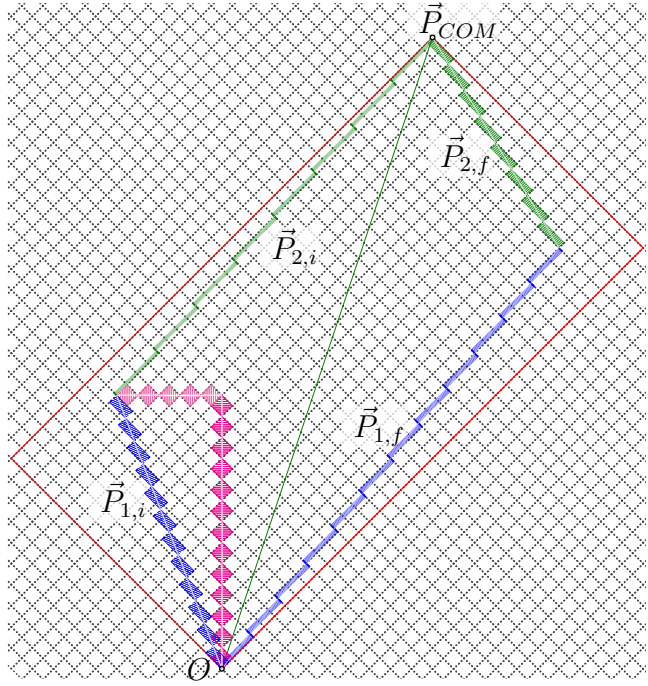 

In addition to kinematical examples in spacetime,
this method can be applied to collision problems in energy-momentum space.
Analogous to the causal diamond of a timelike displacement, let us
consider  
the ``mass diamond'' of a timelike momentum-vector, whose area 
is equal to the square of the associated rest-mass. 
The lengths of momentum-vectors can now be described by tickmarks of mass-units.

In the energy-momentum diagram of Fig.~\ref{fig:EnergyMomentum}, 
we describe an elastic collision%
\cite{fn:elasticCollision}
of
two particles with rest-masses $m_1=12$ and $m_2=8$,
initial-velocities $\VEL_{1,i}=-5/13$ and $\VEL_{2,i}=15/17$,
and final-velocities $\VEL_{1,f}=4/5$ and $\VEL_{2,f}=-3/5$.
From the mass diamond (of size 40 by 20 with Alice's diamonds), 
we have ${k_{\Alice,COM}}^2=(40/20)=2$ so that the 
velocity of the center-of-momentum frame is 
$\VEL\BY{\ALICE}{COM}=(2-1)/(2+1)=1/3$
and the magnitude of the total-momentum 
(called the ``invariant mass'' of the system of particles)
is 
$P_{COM}=\sqrt{(40)(20)}$. 
Alternatively, 
one can construct Alice's right-triangle with 
$P_{COM}$ as its hypotenuse
and determine $\VEL\BY{\ALICE}{COM}=10/30$ and 
$P_{COM}=\sqrt{(30)^2-(10)^2}$.

\section{The Area of a Causal Diamond, Algebraically}
\label{sec:AreaOfACausalDiamond}

In the \textit{geometric construction} of Bob's light-clock diamonds, 
the key feature we exploited was the equality of the areas of 
Alice's and Bob's light-clock diamonds, which is
due to the area-preserving property of the 
Lorentz Boost transformations (Eqs.~(\ref{eq:uTransform}) and (\ref{eq:vTransform})).
However, for the \textit{visualization} of this equality and subsequent
reduction of algebraic computations to counting, 
we implicitly exploited the equality of the areas 
of a figure in a Minkowski spacetime diagram and in a Euclidean space.
In addition, we have not directly addressed the units involved in our construction.
We address these issues now with an algebraic formulation.

Consider Alice's light-clock diamond
(the causal diamond of $OT$ representing 1 tick of Alice's clock).
The area (denoted by $\vec{\cal A}(OT)$) of Alice's diamond  is obtained by 
computing the cross product of its edges
$\VEC{ON}\times\VEC{OM}$.
Using Eq.~(\ref{eq:BasisVectors}), we see that 
$\vec{\cal A}(OT)=\VEC{ON}\times\VEC{OM}=\frac{1}{2} \VEC{OX}\times \VEC{OT}$,
which can be interpreted as half of the cross product of the diamond's diagonals.
Making use of either the Minkowski metric 
(so $\VEC{OT}\cdot \VEC{OT}=1$, $\VEC{OX}\cdot\VEC{OX}=-1$, and $\VEC{OT}\cdot\VEC{OT}=1$) 
or the usual Euclidean metric, we determine that
$\frac{1}{2}\VEC{OX}\times \VEC{OT}$ has ``size'' $\frac{1}{2}\rm{\ tick}^2$. 
(Although square-norms of this cross product may disagree in sign 
between the two metrics [depending on signature conventions],
they fortunately agree in absolute value.)
Since this factor of $\frac{1}{2}$ will complicate our 
\textit{counting} calculations,
we have chosen the units of diamond-area to be ``the area of Alice's light-clock diamond,'' 
or simply ``diamond'' (as we have done earlier), 
rather than $\rm{tick}^2$. 

In addition,
since $\vec{\cal A}$ carries the units of area,
the coordinates $(t,x)$ and $(u,v)$ must be dimensionless, as we declared in
Sec.~\ref{sec:coordsys}. (Had we written 
$t_T=1\rm{\ tick}$ and $x_T=0\rm{\ ticks}$,
so that
$u_T=1\rm{\ tick}$ and $v_T=1\rm{\ tick}$,
we would have $u_T v_T=1\rm{\ tick}^2$ for the ``area.''
However, we just found that
the area of Alice's diamond is only $\frac{1}{2}\rm{\ tick}^2$.)

To incorporate all of these issues, 
we will perform an algebraic computation of the square-interval of $OQ$ in
Figs.~\ref{fig:AliceCoords} and~\ref{fig:AliceLightConeCoords} as an area.
Recall Eq.~(\ref{eq:OQvector}):
\begin{eqnarray*}
     \VEC{OQ} &=& t_Q \VEC{OT} + x_Q \VEC{OX} \ =\ u_Q \VEC{ON} + v_Q \VEC{OM}.
\end{eqnarray*}
We now show that $\vec{\cal A}(OQ)$, the area of the causal diamond of $OQ$, is
equal to the square-interval of $OQ$ in units of $\vec{\cal A}(OT)$, the area of Alice's diamond.
Starting with the edges of the causal diamond, we have
\begin{eqnarray}
 \vec{\cal A}(OQ) 
&\stackrel{?}{=}&
\vec{\cal A}(OT)
(\VEC{OQ} \cdot \VEC{OQ})\  
 \nonumber
 \\
u_Q \VEC{ON} \times v_Q \VEC{OM} 
&\stackrel{?}{=}&
\vec{\cal A}(OT)
(t_Q \VEC{OT} + x_Q \VEC{OX}) \cdot 
(t_Q \VEC{OT} + x_Q \VEC{OX})\ 
\nonumber
\\
u_Q v_Q\ \VEC{ON} \times \VEC{OM} 
&\stackrel{?}{=}&
\vec{\cal A}(OT)
({t_Q}^2\VEC{OT}\cdot\VEC{OT} +
{x_Q}^2\VEC{OX}\cdot\VEC{OX} 
+2{t_Q}{x_Q}\VEC{OT}\cdot\VEC{OX})
\nonumber
\\
\left({t_Q}+{x_Q}\right) \left({t_Q}-{x_Q}\right) \ \vec{\cal A}(OT) 
&\stackrel{\surd}{=}&
\vec{\cal A}(OT)({t_Q}^2- {x_Q}^2)
\label{eq:AreaCalculation}
\end{eqnarray}
where we have used Eq.~(\ref{eq:LightConeCoords}) and the Minkowski metric.
Alternatively, we can start with the diagonals of the causal diamond, 
$\VEC{OQ}$ and 
$\VEC{OQ}_\perp=u_Q \VEC{ON} - v_Q \VEC{OM}$,
then compute $\vec{\cal A}(OT)=\frac{1}{2}\VEC{OQ}_\perp \times \VEC{OQ}$
to arrive at the same result.
By applying Eqs.~(\ref{eq:uTransform}) and (\ref{eq:vTransform}),
the invariance under a Lorentz Boost transformation is easily seen.

\section{Summary}

We have shown how calculations in Special Relativity are 
facilitated by visualizing ticks of a clock
as light-clock diamonds drawn
on a sheet of rotated graph paper.
When the relative-velocities between observers 
have rational Bondi-Doppler $k$-factors,
the arithmetic and graphical constructions become very simple.
This allows us to place emphasis first on the
physical interpretation and geometrical modeling 
of situations in special relativity. 
Once established, one can then (if desired) 
advance to the development of the standard relativistic formulas.

\begin{acknowledgments}

I wish to thank my students in AST 110L and PHY 216 at Mount Holyoke College
and in PHYS 63 at Bowdoin College, who tried out various worksheets based on
this work. Their performance and feedback helped improve the presentation of the material.
I also wish to thank Tevian Dray, Tom Moore, Stephen Naculich 
for useful comments.

\end{acknowledgments}

\appendix*
\section{Formulae for Velocity Composition and the Lorentz Boost Tranformation}
\label{sec:appendix}

For completeness, we show how one can obtain the formulae
for Velocity Composition and the 
Lorentz Boost coordinate transformation 
using Radar Measurements. 
We follow the methods of Bondi\cite{Bondi} and Ellis.\cite{EllisWilliams}
The starting point for both formulae begins with
\begin{eqnarray}
t\BY{\ALICE}{R} &=& k_{\Alice,\Bob}\ t\BY{\BOB}{S}\label{eq:RkS}\\
t\BY{\BOB}{\vphantom{S}s} &=&k_{\Alice,\Bob}\ t\BY{\ALICE}{\vphantom{R}r},\label{eq:skr}
\end{eqnarray}
from Fig.~\ref{fig:RadarMeasurements} and Eq.~(\ref{eq:BondiK}). 

We first obtain the Velocity Composition formula, Eq.~(\ref{eq:velcomp}).
From  Sec.~\ref{sec:Radar},
$k_{\Alice,\Carol}/k_{\Bob,\Carol}=\sqrt{  ( 
t\BY{\ALICE}{R} / t\BY{\BOB}{S}) 
(t\BY{\BOB}{\vphantom{S}s}/ t\BY{\ALICE}{\vphantom{R}r})}=k_{\Alice,\Bob}$,
since each term in parenthesis is equal to $k_{\Alice,\Bob}$.
[Alternatively, write Eq.~(\ref{eq:RkS}) as 
$t\BY{\ALICE}{R} = k_{\Alice,\Bob}\ t\BY{\BOB}{S}
=k_{\Alice,\Bob} (k_{\Bob,\Carol}\ t\BY{\CAROL}{W})=k_{\Alice,\Carol}\ t\BY{\CAROL}{W}$,
using Fig.~\ref{fig:RadarMeasurements} and Eq.~(\ref{eq:BondiK}) again.]
Thus,
\begin{equation}
k_{\Alice,\Carol}=k_{\Alice,\Bob}\ k_{\Bob,\Carol},
\label{eq:velcompK}
\end{equation}
which embodies velocity composition.
(Proof: Plug this expression into Eq.~(\ref{eq:vAC}), then use Eq.~(\ref{eq:kvel})
twice to obtain Eq.~(\ref{eq:velcomp}).)

We now obtain the %
Lorentz Boost coordinate transformation formulae, Eqs.~(\ref{eq:tBoost}) and (\ref{eq:xBoost}).
(It may help to recall Eqs.~(\ref{eq:rectCoordsA}) and~(\ref{eq:rectCoordsB}).)
Using Eqs.~(\ref{eq:RkS}) and (\ref{eq:skr}) with $k=k_{\Alice,\Bob}$, 
write 
\begin{align}
\frac{1}{2}\left(t\BY{\ALICE}{R} \pm t\BY{\ALICE}{\vphantom{R}r}\right)
&= \frac{1}{2}\left(
k\ t\BY{\BOB}{S} \pm k^{-1}\ t\BY{\BOB}{\vphantom{S}s} 
\right)\nonumber\\
&= \frac{1}{2}\left( 
k \left( t\BY{\BOB}{W} + x\BY{\BOB}{W} \right) \pm k^{-1} \left( t\BY{\BOB}{W} - x\BY{\BOB}{W} \right)
\right)\nonumber\\
&= \left(\frac{k \pm k^{-1}}{2}\right)
 t\BY{\BOB}{W} +  \left(\frac{k \mp  k^{-1}}{2}\right) x\BY{\BOB}{W} \label{eq:BoostFormula0},
\end{align}
where we used Eq.~(\ref{eq:rectCoordsB}) to introduce Bob's rectangular coordinates of event $W$.
Using Eq.~(\ref{eq:kvel}), we have $(k+k^{-1})/2=1/\sqrt{1-\beta^2}=\gamma$ and 
$(k-k^{-1})/2=\beta/\sqrt{1-\beta^2}=\beta\gamma$.
With the upper signs in Eq.~(\ref{eq:BoostFormula0}), 
the left-hand side is $t\BY{\ALICE}{W}$ (using Eq.(\ref{eq:rectCoordsA})), 
and we obtain the analogue of
Eq.~(\ref{eq:tBoost}).
With the lower signs, 
the left-hand side is $x\BY{\ALICE}{W}$, and we obtain the analogue of
Eq.~(\ref{eq:xBoost}).


\begin{thebibliography}{99}
\bibitem{fn:area}
The magnitude of the ordinary Euclidean area is equal 
to the magnitude of the area in Minkowski spacetime,
up to a constant scalar factor. 
See Sec.~\ref{sec:AreaOfACausalDiamond} and Ref.~\onlinecite{Jacobson}.
\bibitem{fn:conventions}
We use units where temporal displacements are measured in
units of ``ticks'' and spatial displacements are also measured in
ticks, the times needed by light to travel those spatial displacements.
In these units, velocities are dimensionless, and the speed of light is equal to one. 
In addition, we use the timelike-signature convention
so that the square-interval is $ds^2=dt^2-dx^2$.
We follow the usual convention where time runs upwards in the diagram.
\bibitem{fn:MinkowskiNormal}
This is equivalent to Minkowski's 
definition: the radius vector drawn to a point on the
hyperbola is perpendicular to the tangent vector at that point.
(See Ref.~\onlinecite{Minkowski}, p.~85.)
\bibitem{fn:areaCalc}
This notion of ``area'' will be more fully explained in 
Sec.~\ref{sec:AreaOfACausalDiamond}.
\bibitem{fn:decodeTransmissions}
We assume that Alice and Bob can suitably encode and decode the transmissions.
\bibitem{fn:perpendicular}
Indeed, this is one of the properties that 
makes Minkowski spacetime diagrams
difficult to interpret. 
Note, however, that Galilean spacetime diagrams 
[ordinary position-time graphs] also have this property.
\bibitem{fn:PythagoreanTriples}
Since $(\Delta x)^2+(\Delta s)^2=(\Delta t)^2$ 
for nonnegative integers $\Delta x$, $\Delta s$, and $\Delta t$, 
with $\Delta x < \Delta t$ and $\Delta s \leq \Delta t$ for future-timelike displacements,
then
$(\Delta x , \Delta s , \Delta t)$ form a Pythagorean triple.
When such triples are generated by
$\Delta x=\lambda( m^2-n^2 )$,
$\Delta s=\lambda( 2mn )$, and
$\Delta t=\lambda( m^2+n^2)$,
with positive integers $\lambda$, $m$, and $n$ (with $m\geq n$),
we find that $k=m/n$.
Alternatively, when the generator-formulas for $\Delta x$ and $\Delta s$ are switched
(i.e., $\Delta s=\lambda( m^2-n^2 )$ and $\Delta x=\lambda( 2mn )$),
we find that $k=(m+n)/(m-n)$.
\bibitem{fn:elasticCollision}
Based on Fig 7.6 (p. 207) in Taylor and Wheeler (Ref.~\onlinecite{TaylorWheelerB}).
Note that the area of the mass-diamond or the tickmarks of mass-units
can now replace the ``handles'' on the momentum-vectors, first described on p. 198.
See also problem R10.S3 (p. 189) in Moore (Ref.~\onlinecite{Moore}).
\end{thebibliography}
\end{document}